\documentclass[cmp]{svjour}

\usepackage{amsmath,amsfonts,amssymb,times}
\usepackage{graphicx}

\newcommand{\trace}{\mathop{\rm Tr}\nolimits}

\newcommand{\diag}{\mathop{\rm Diag}\nolimits}

\newcommand{\supp}{\mathop{\rm Supp}\nolimits}

\newcommand{\bra}[1]{\langle#1|}
\newcommand{\ket}[1]{|#1\rangle}

\newcommand{\twomat}[4]{{\left(\begin{array}{cc}#1&#2\\#3&#4\end{array}\right)}}

\newcommand{\R}{{\mathbb{R}}}

\DeclareRobustCommand\openone{\leavevmode\hbox{\small1\normalsize\kern-.33em1}}
\newcommand{\id}{\mathrm{\openone}}

\newcommand{\dds}{\frac{\partial}{\partial s}}

\newcommand{\be}{\begin{equation}}
\newcommand{\ee}{\end{equation}}
\newcommand{\bea}{\begin{eqnarray}}
\newcommand{\eea}{\end{eqnarray}}
\newcommand{\beas}{\begin{eqnarray*}}
\newcommand{\eeas}{\end{eqnarray*}}

\begin{document}
\title{Asymptotic Error Rates in Quantum Hypothesis Testing}
\author{K.M.R.~Audenaert\inst{1,2} \and M.~Nussbaum\inst{3} \and A.~Szko{\l}a\inst{4} \and F.~Verstraete\inst{5}}
\institute{
Institute for Mathematical Sciences, Imperial College London, 53 Prince's Gate, London SW7 2PG, UK,\\
\email{k.audenaert@imperial.ac.uk}
\and
Dept.\ of Mathematics, Royal Holloway, University of London, Egham, Surrey TW20 0EX, UK
\and
Department of Mathematics, Cornell University, Ithaca, NY 14853, USA,\\
\email{nussbaum@math.cornell.edu}
\and
Max Planck Institute for Mathematics in the Sciences,
Inselstrasse 22, 04103 Leipzig, Germany,\\
\email{szkola@mis.mpg.de}
\and
Fakult\"at f\"ur Physik, Universit\"at Wien, Boltzmanngasse 5, 1090 Wien, Austria,\\
\email{frank.verstraete@univie.ac.at}
}
\date{\today}
%

\def\makeheadbox{}
\maketitle

\begin{abstract}
We consider the problem of discriminating between two different states of a finite quantum system in the setting of large numbers of
copies, and find a closed form expression for the asymp\-totic exponential rate at which the specified error probability tends to zero.
This leads to the identification of the quantum generalisation of the classical Chernoff distance, which is the
corresponding quantity in classical symmetric hypothesis testing,
thereby solving a long standing open problem.

The proof relies on a new trace inequality for pairs of positive operators as well as on a special mapping
from pairs of density operators to pairs of probability distributions.
These two new techniques have been introduced in [quant-ph/0610027] and [quant-ph/0607216], respectively.
They are also well suited to prove the quantum generalisation of the Hoeffding bound, which
is a modification of the Chernoff distance and specifies the optimal achievable asymptotic error rate in
the context of asymmetric hypothesis testing. This has been done subsequently by  Hayashi [quant-ph/0611013]
and Nagaoka [quant-ph/0611289] for the special case where both hypotheses have full support.

Moreover, quantum Stein's Lemma and quantum Sanov's theorem may be derived directly from quantum Hoeffding
bound combining it with a result obtained recently in [math/0703772].

The goal of this paper is to present the proofs of the above mentioned results in a unified way
and in full generality (allowing hypothetic states with different supports) using mainly the techniques
from [quant-ph/0607216] and [quant-ph/0610027].

Additionally, we give an in-depth treatment of the properties of the quantum Chernoff distance.
We argue that, although it is not a metric, it is a
natural distance measure on the set of density operators, due to its clear operational meaning.
\end{abstract}
%
\section{Introduction}\label{sec:intro}
One of the basic tasks in information theory is discriminating
between two different information sources, modelled by (time-discrete) stochastic processes.
Given a source that generates independent, identically distributed (i.i.d.) random variables,
according to one out of two possible probability distributions,
the task is to determine which distribution is the true one, and to do so with minimal error,
whatever error criterion one chooses.

This basic decision problem has an equally basic quantum-informational incarnation.
Given an information source that emits quantum systems (particles) independently and identically
prepared in one out of two possible quantum states,
figure out which state is the true one, with minimal error probability.

In both settings, we're dealing with two hypotheses, each one pertaining to one law represented
by a probability distribution or a quantum state, respectively,
and the discrimination problem is thus a particular instance of a hypothesis testing problem.

In hypothesis testing, one considers a null hypothesis and an alternative hypothesis.
The alternative hypothesis is the one of interest and states that ``something significant is happening'',
for example, a cell culture under investigation is coming from a malignant tumor, or some case of flu
is the avian one, or an e-mail attachment is a computer virus.
In contrast, the null hypothesis corresponds to this not being the case; the cells are normal ones,
the flu can be treated with an aspirin, and the attachment is just a nice picture.
This is inherently an asymmetric situation,
and Neyman and Pearson introduced the idea of similarly making a distinction between type I and type II errors.
\begin{itemize}
\item The type I error or ``false positive'', denoted by $\alpha$, is the error of accepting the alternative hypothesis
when in reality the null hypothesis holds and the results can be attributed merely to chance.
\item The type II error or ``false negative'', denoted by $\beta$, is the error of accepting the null hypothesis
when the alternative hypothesis is the true state of nature.
\end{itemize}
The costs associated to the two types of error can be widely different, or even incommensurate.
For example, in medical diagnosis, the type I error corresponds to diagnosing a healthy patient with a certain affliction,
which can be an expensive mistake, causing a lot of grievance.
On the other hand, the type II error may correspond to declaring a patient healthy while in reality
(s)he has a life-threatening condition, which can be a fatal mistake.

To treat the state discrimination problem as a hypothesis test,
we assign the null hypothesis to one of the two states
and the alternative hypothesis to the other one.
If all we want to know is which one of the two possible states we are observing,
the mathematical treatment is completely symmetric under the interchange of these two states.
It therefore fits most naturally in the setting
of \textit{symmetric hypothesis testing}, where no essential distinction is made between the two kinds of errors.
To wit, in symmetric hypothesis testing, one considers the average, or Bayesian, error probability $P_e$, defined
as the average of $\alpha$ and $\beta$
weighted by the prior probabilities of the null and the alternative hypothesis, respectively.

This paper will be concerned with symmetric as well as with asymmetric quantum hypothesis testing.
Since we have developed the main techniques in the symmetric setting we will start with this case and
address the asymmetric setting at the end.

The optimal solution to the symmetric classical hypothesis test is given by
the maximum-likelihood (ML) test. Starting from the outcomes of an experiment
involving $n$ independent draws from the unknown distribution,
one calculates the conditional probabilities
(likelihoods) that these outcomes can be obtained when the distribution is the one of the null hypothesis
and the one of the alternative hypothesis, respectively.
One decides then on the hypothesis for which the conditional probability is the highest.
I.e.\ if the \textit{likelihood ratio} is higher than 1, the null hypothesis is rejected, otherwise it is accepted.

In the quantum setting, the experiment consists of preparing $n$ independent copies of a quantum system in an
unknown state, which is either $\rho$ or $\sigma$,
and performing an optimal measurement on them. We assume that the quantum systems are finite, implying that
the states are associated to density operators on a finite-dimensional complex Hilbert space.
Under the null hypothesis, the combined $n$ copies correspond to an $n$-fold tensor product density operator $\rho^{\otimes n}$, while
under the alternative hypothesis, the associated density operator is $\sigma^{\otimes n}$.
The null hypothesis is then accepted or rejected according to the outcome of the measurement and the specified decision rule.
The task of finding this optimal measurement
is so fundamental that it was one of the first problems considered in the field of quantum
information theory; it was solved in the one-copy case more than 30 years ago by Helstrom and Holevo \cite{helstrom,holevo}.
We refer to the generalised ML-tests as Holevo-Helstrom tests.
In the special case of equal priors, the associated minimal probability of error achieved by the optimal measurement
can be calculated from the trace norm distance between the two states:
\begin{eqnarray}\label{def:P^*}
P_{e,n}^*(\rho, \sigma) = \frac{1}{2}(1-\|\rho^{\otimes n}-\sigma^{\otimes n}\|_1/2),
\end{eqnarray}
where $\|A\|_1:=\trace|A|$ denotes the trace norm.

Going back to the classical case again, in a seminal paper, H.~Chernoff \cite{chernoff}
investigated the so-called \textit{asymptotical efficiency} of a class of statistical tests,
which includes the likelihood ratio test mentioned before.
The probability of error $P_{e,n}$ in discriminating two probability distributions
decreases exponentially in $n$, the number of draws from the distribution: $P_{e,n}\sim \exp(-\xi n)$.
For finite $n$ this is a rather crude approximation. However, as $n$ grows larger
one finds better and better agreement, and the exponent $\xi$ becomes meaningful in
the asymptotic limit. The asymptotical efficiency is exactly the asymptotic limit of this exponent.

Chernoff was able to derive an (almost) closed expression for this asymptotic efficiency,
which was later named eponymously in his honour.
For two discrete probability distributions $p$ and $q$, this expression is given by
\be\label{eq:cbcl}
\xi_{CB}(p,q):=-\log\left( \inf_{0\le s\le 1} \sum_i p(i)^{1-s} q(i)^{s}\right),
\ee
which is of closed form but for a single variable minimisation.
This quantity goes under the alternative names of Chernoff distance, Chernoff divergence and Chernoff information.

While Chernoff's main purpose was to use this asymptotic efficiency measure to compare the power of different tests
-- the mathematically optimal test need not always be the most practical one --
it can also be used as a distinguishability measure between the distributions (states) of the
two hypotheses. Indeed, fixing the test, its efficiency for a particular pair of distributions
gives a meaningful indication of how well these two distributions can be distinguished by that test.
This is especially meaningful if the applied test is the optimal one.

A quantum generalisation of Chernoff's result is highly desirable. Given the large amount of experimental effort
in the context of quantum information processing to prepare and measure quantum states, it is of fundamental
importance to have a theory that allows to discriminate different quantum states in a meaningful way. Despite considerable effort,
however, the quantum generalisation of the Chernoff distance has until recently remained unsolved.

In the previous papers, \cite{szkola} and \cite{spain}, this issue was finaly settled
and the asymptotic error exponent was identified, when the optimal Holevo-Helstrom
strategy for discriminating between the two states is used, by
proving that the following version of the Chernoff distance
\be\label{eq:qcb}
\xi_{QCB}(\rho, \sigma):=-\log\left( \inf_{0\le s\le 1} \trace[\rho^{1-s} \sigma^{s}]\right),
\ee
  has the same operational meaning as its classical counterpart: It specifies the asymptotic rate
  exponent of the minimal error probability $P_{e,n}^*$ (recall definition (\ref{def:P^*})).
  Remarkably, it looks like an almost na{\"\i}ve generalisation of the classical expression (\ref{eq:cbcl}).

We remark that in the literature different extensions of the classical expression have been considered.
Indeed, when insisting only on the compatibility with the classical Chernoff distance, there is in principle an infinitude of possiblities.
Among those, three especially promising candidate expressions had been put forward by Ogawa and Hayashi \cite{hayashi},
who studied their relations and found that there exists an increasing ordering between them.
Incidentally, the second candidate coincides with (\ref{eq:qcb}) and thus turns out to be the correct one.

Kargin \cite{kargin} gave lower and upper bounds on the optimal error
exponent  $\xi$ in terms of the fidelity between the two density operators
and found that Ogawa and Hayashi's third candidate (in their increasing
arrangement) is a lower bound on the optimal error exponent for faithful
states, i.e.\ it is an achievable rate. Hayashi
\cite{hayashibook} made progress regarding (\ref{eq:qcb}), by showing that
for
$s=1/2$, $ - \log \trace[\rho^{1-s} \sigma^{s}]$
is also an achievable error exponent.

The proof of our main result consists of two parts. In the optimality part, which was first
presented in \cite{szkola}, we show that for any test the (Bayesian) error rate
$-\frac{1}{n}\log P_{e,n}$ cannot be made arbitrary large but is asymptotically
bounded above by $\xi_{QCB}$.
In the achievability part, first put forward in \cite{spain}, we prove that under the Holevo-Helstrom
strategy the bound is actually attained in the asymptotic limit,
i.e.\ \[\limsup_{n \to \infty}\left(-\frac{1}{n}\log P^{\ast}_{e,n}\right)\geq \xi_{QCB}.\]

It is the purpose of this paper to give a complete, detailed, and unified account of these results.
We will present the complete proof in Section \ref{sec:chernoff}.
Moreover, we give an in-depth treatment of the properties of the quantum Chernoff distance in Section \ref{sec:properties}.
More precisely, we show that it defines a distance measure between quantum states.

Distinguishability measures between quantum states have been used in a wide variety of applications in quantum information
theory. The most popular of such measures seems to be Uhlmann's fidelity \cite{Uhlman}, which happens to coincide with the quantum
Chernoff distance when one of the states is pure. The trace norm distance $\|\rho-\sigma\|_1 = \trace|\rho-\sigma|$
has a more natural operational meaning than the fidelity, but
lacks monotonicity under taking tensor powers of its arguments. The problem is that one can easily find states
$\rho,\sigma,\rho',\sigma'$ such that $\|\rho-\sigma\|_1 < \|\rho'-\sigma'\|_1$ but $\|\rho^{\otimes
2}-\sigma^{\otimes 2}\|_1 > \|\rho'^{\otimes 2}-\sigma'^{\otimes 2}\|_1$.
This already happens in the classical setting: take the following 2-dimensional diagonal states
$$
\rho=\twomat{1/4}{0}{0}{3/4},\sigma=\twomat{3/4}{0}{0}{1/4},\rho'=\twomat{0}{0}{0}{1},\sigma'=\twomat{b}{0}{0}{1-b},
$$
where $1-1/\sqrt{2}< b<1/2$.
Then $\|\rho-\sigma\|_1=1>2b=\|\rho'-\sigma'\|_1$, while
$\|\rho^{\otimes 2}-\sigma^{\otimes 2}\|_1 = 1<2b(2-b)=\|\rho'^{\otimes 2}-\sigma'^{\otimes 2}\|_1$.
The quantum Chernoff distance
characterises the exponent arising in the asymptotic behaviour of the trace norm distance, in the case of many identical
copies, and therefore by construction does not suffer from this problem. As such, the quantum Chernoff distance can be
considered as a kind of regularisation of the trace norm distance.
For the above-mentioned states, $\xi_{QCB}(\rho,\sigma)=-\log(\sqrt{3}/2)$ (optimal $s=1/2$)
and $\xi_{QCB}(\rho',\sigma')=-\log(1-b)$ (optimal $s=1$).

A related problem that attracted a lot of attention in the field of quantum information theory was to identify the
relative entropy between two quantum states. An information-theoretical way of looking at the classical relative entropy
between two probability distributions, or Kullback-Leibler distance, is that it characterises the inefficiency of
compressing messages from a source $p$ using an algorithm that is optimal for a source $p'$
(i.e.\ yields the Shannon information bound for that source). Phrased differently, it quantifies the way one could
cheat by telling that the given probability distribution is $p$ while the real one is $p'$. By proving a quantum
version of Stein's lemma \cite{hiaipetz,ogawa}, it has been shown that the quantum relative entropy,
as introduced by Umegaki, has exactly the same operational meaning.

When using the relative entropy to distinguish between states, one faces the problem that it is not continuous and is asymmetric
under exchange of its arguments, and therefore it does not represent a distance measure in mathematically strict manner.
Furthermore, for pure states, the quantum relative entropy is not very useful, since it is either 0 (when the
two states are identical) or infinite (when they are not).
In contrast, the quantum Chernoff distance seems to be much more natural in many situations.

On the other hand, (quantum) relative entropy is a crucial notion in \textit{asymmetric} hypothesis testing.
There it obtains an operational meaning as the best achievable asymptotic rate of type II errors. Its properties,
which are problematic for a candidate for a distance measure, reflect the asymmetry between the null and alternative
hypothesis arising from treating the type-I and type-II errors in a different way.
As exemplified by the medical diagnosis case mentioned above, the type II error is the one that should
be avoided at all costs.
Hence, one puts a constraint $\alpha < \epsilon$ on the type I error, and minimises the $\beta$-rate.
One obtains that the optimal $\beta$-rate is the relative entropy of the null hypothesis w.r.t.\ the alternative,
independent of the constrained $\epsilon$. The mathematical derivation of this statement goes under the name of Stein's Lemma.
When the constraint consists of a lower bound on the asymptotic exponential rate of the type II error,
one obtains what is called the Hoeffding bound.

Asymmetric hypothesis testing has been subject to a quantum theoretical treatment much earlier, although it
is a much less natural setting for the basic state discrimination problem.
The quantum generalisation of Stein's Lemma was first obtained by Hiai and Petz \cite{hiaipetz}.
Its optimality part was then strengthened by Ogawa and Nagaoka in \cite{ogawa}.
In the last few years there has been a lot of progress extending the statement of the lemma in different directions.
In  \cite{qSanov} the minimal relative entropy distance from a set of quantum states, the null hypothesis,
w.r.t.\ a reference quantum state, the alternative, has been fixed as the best achievable asymptotic rate of
the type II errors, see also \cite{hayashi:universal_stein}. This may be seen as a quantum generalisation of Sanov's theorem.
In a recent paper \cite{qSanov2} an extension of this result to the case where the hypotheses correspond
to sources emitting correlated (not necessarily i.i.d.) classical or quantum data has been given.
Additionally, an equivalence relation between the achievability part in (quantum) Stein's Lemma and (quantum) Sanov's Theorem has been derived.

Just a few months after the appearance of \cite{szkola,spain}, the techniques pioneered in those two papers
were used to find a quantum generalisation
of the Hoeffding bound under the implicit assumption of equivalent hypotheses, i.e.\ for states with coinciding supports,
thereby (partially) solving another long-standing open problem in quantum hypothesis testing.
Just as in the case of the Chernoff distance, the Hoeffding bound contains $\sum_i p(i)^{1-s} q(i)^{s}$
as a sub-expression, and the quantum generalisation of the Hoeffding bound is obtained
by replacing this sub-expression by $\trace[\rho^{1-s} \sigma^{s}]$.
The optimality of the bound (also called the ``converse part'') was proven by Nagaoka \cite{nagaoka06},
while its achievability (the ``direct part'') was found by Hayashi \cite{hayashi06}.
Using the same techniques, Hayashi also gave a simple proof of the achievability part of the
quantum Stein's Lemma, in that same paper. In Section \ref{sec:hoeffding} we first formulate and prove an extended
version of the classical Hoeffding bound, which allows nonequivalent hypotheses. Secondly, we present a complete
proof of the quantum Hoeffding bound in a unified way. Moreover, we derive quantum Stein's Lemma as well as quantum
Sanov's Theorem from the quantum Hoeffding bound combined with the mentioned equivalence relation proved in \cite{qSanov2}.

%
\section{Mathematical Setting and Problem Formulation}\label{sec:problem_formulation}

We consider the two hypotheses $H_0$ (null) and $H_1$ (alternative) that a device
 prepares finite quantum systems either in the state $\rho$
or in the state $\sigma$, respectively.
Everywhere in this paper, we identify a state with a density operator, i.e.\ a positive trace $1$ linear operator
on a finite-dimensional Hilbert space $\mathcal{H}$ associated to the type of the finite quantum system in question.
Since the (quantum) Chernoff distance arises naturally in a Bayesian setting, we
supply the prior probabilities $\pi_0$ and $\pi_1$, which are positive quantities summing up to 1;
we exclude the degenerate cases $\pi_0=0$ and $\pi_1=0$ because these are trivial.

Physically discriminating between the two hypotheses corresponds to performing a generalised (POVM) measurement on the
quantum system. In analogy to the classical proceeding one accepts $H_0$ or $H_1$ according to a decision rule based
on the outcome of the measurement. There is no loss of generality assuming that the POVM consists of only two elements, which we denote by
$\{\id-\Pi,\Pi\}$, where $\Pi $ may be any linear operator on $\mathcal{H}$ with $0\le \Pi\le \id$.
We will mostly make reference to this POVM by its $\Pi$ element, the one corresponding to the alternative hypothesis.
The type-I and type-II error probabilities $\alpha$ and $\beta$ are the probabilities of
mistaking $\sigma$ for $\rho$, and vice-versa, and are given by
\beas
\alpha &:=& \trace[\Pi\rho] \\
\beta  &:=& \trace[(\id-\Pi)\sigma].
\eeas
The average error probability $P_e$ is given by
\be\label{eq:pe}
P_e=\pi_0\alpha+\pi_1\beta=\pi_0\trace[\Pi\rho] + \pi_1\trace[(\id-\Pi)\sigma].
\ee
The Bayesian distinguishability problem consists in finding the $\Pi$ that minimises $P_e$.
A special case is the symmetric one where the prior probabilities $\pi_0, \pi_1$ are equal.

\bigskip

Before we proceed, let us first introduce some basic notations. Abusing terminology, we will use the term `positive' for
`positive semi-definite' (denoted $A\ge0$). We employ the positive semi-definite ordering on the linear operators on $\mathcal{H}$
throughout, i.e.\ $A\ge B$ iff $A-B\ge0$. For each linear operator $A \in \mathcal{B}(\mathcal{H})$
the \textit{absolute value} $|A|$ is defined as $|A|:=(A^* A)^{1/2}$. The Jordan
decomposition of a self-adjoint operator $A$ is given by $A=A_+ - A_-$, where
\be\label{eq:posdef}
A_+:=(|A|+A)/2,\qquad A_-:=(|A|-A)/2
\ee
are the
\textit{positive part} and \textit{negative part} of $A$, respectively.
Both parts are positive by definition, and $A_+A_-=0$.

There is a very useful variational characterisation of the trace of the positive part of a self-adjoint operator $A$:
\be\label{eq:trpos1}
\trace[A_+] = \max_X \{\trace[AX]: 0\le X\le \id\}.
\ee
In other words, the maximum is taken over all positive contractive operators.
Since the extremal points of the set of positive contractive operators are exactly the
orthogonal projectors, we also have
\be\label{eq:trpos2}
\trace[A_+] = \max_P \{\trace[AP]: P\ge0, P=P^2\}.
\ee
The maximiser on the right-hand side is the orthogonal projector onto the range of $A_+$.

\bigskip

We can now easily prove the quantum version of the Neyman-Pearson Lemma.
\begin{lemma}[Quantum Neyman-Pearson]\label{th:neyman-pearson}
Let $\rho$ and $\sigma$ be density operators associated to hypotheses $H_0$ and $H_1$, respectively.
Let $T$ be a fixed positive number. Consider the POVM with elements $\{\id-\Pi^*,\Pi^*\}$ where $\Pi^*$ is
the projector onto the range of $(T\sigma-\rho)_+$, and let $\alpha^*=\trace[\Pi^*\rho]$ and $\beta^*=\trace[(\id-\Pi^*)\sigma]$
be the associated errors. For any other POVM $\{\id-\Pi,\Pi\}$, with associated errors $\alpha=\trace[\Pi\rho]$
and $\beta=\trace[(\id-\Pi)\sigma]$, we have
$$
\alpha+T\beta \ge \alpha^*+T\beta^* = T-\trace[(T\sigma-\rho)_+].
$$
Thus if $\alpha\le\alpha^*$, then $\beta\ge\beta^*$.
\end{lemma}
\textit{Proof.}
By formulae (\ref{eq:trpos1}) and (\ref{eq:trpos2}),  for all $0 \leq \Pi \leq \id$ we have
$\trace[\Pi(T\sigma-\rho)] \leq \trace(T\sigma-\rho)_+ = \trace[\Pi^*(T\sigma-\rho)] $.
In terms of $\alpha,\beta,\alpha^*,\beta^*$, this reads
$T(1-\beta)-\alpha \leq T(1-\beta^*)-\alpha^*$,
which is equivalent to the statement of the Lemma.
\qed

\medskip

The upshot of this Lemma is that the POVM $\{\id-\Pi^*,\Pi^*\}$, where $\Pi^*$ is
the projector on the range of $(T\sigma-\rho)_+$, is the optimal one when the goal is to
minimise the quantity $\alpha+T\beta$. In symmetric hypothesis testing the positive number $T$ is taken to be the ratio
$\pi_1 / \pi_0$ of the prior probabilities.

\medskip

We emphasize that we have started with the assumption that the physical systems in question
are finite systems with an algebra of observables $\mathcal{B}(\mathcal{H})$,
i.e.\ the algebra of linear operators on a finite-dimensional Hilbert space $\mathcal{H}$. This is a purely quantum situation.
In the general setting (of statistical mechanics) one associates to a finite physical system, classical or quantum,
a finite-dimensional $\ast$-algebra $\mathcal{A}$. Such an algebra has a block representation
$\bigoplus_{i=1}^k \mathcal{B}(\mathcal{H}_i)$, i.e.\ it is a subalgebra of $\mathcal{B} (\mathcal{H})$,
where $\mathcal{H}:=\bigoplus_{i=1}^k \mathcal{H}_i $. If the Hilbert spaces $\mathcal{H}_i$ are one-dimensional for
all $i=1, \ldots,k$, then $\mathcal{A}$ is $^{\ast}$-isomorphic to the commutative algebra of diagonal $(k \times k)$-matrices.
This covers the classical case.  Now, in view of Lemma \ref{th:neyman-pearson} it becomes
clear that in the context of hypothesis testing there is no restriction assuming that the algebra of observables of the systems in question
is $\mathcal{B}(\mathcal{H})$;
indeed, the optimally discriminating projectors $\Pi^*$  are always in the $\ast$-subalgebra generated by the two involved density
operators $\rho$ and $\sigma$. This implies that they are automatically elements of the algebra  $\mathcal{A}$ characterising
the physical systems. In particular, if the hypotheses correspond to mutually commuting density operators then the problem reduces
to a classical one in the sense that the best test  $\Pi^*$ commutes with the density operators as well.
Hence it coincides with the classical ML-test, although there are many more possible tests in  $\mathcal{B}(\mathcal{H})$ than
in the commutative subalgebra of observables of the  classical subsystem.

\medskip

The basic problem we focus on in this paper is to identify how the error probability
$P_e$ behaves in the asymptotic limit, i.e.\ when one has to discriminate between
the hypotheses $H_0$ and $H_1$ on the basis of a large number $n $ of copies of the quantum systems. This means that we have
to distinguish between the $n$-fold tensor product density operators $\rho^{\otimes n}$ and $\sigma^{\otimes n}$ by means of
POVMs $\{\id-\Pi_n,\Pi_n\}$ on $\mathcal{H}^{\otimes n}$.

We define the rate limit $s_R$ for any positive sequence  $(s_n)$ as
$$
s_R:=\lim_{n\to\infty} \left( -\frac{1}{n} \log s_n \right),
$$
if the limit exists. Otherwise we have to deal with the lower and upper rate limits $\underline{s}_R$ and $\overline{s}_R$,
which are the limit inferior and the limit superior of the sequence $(-\frac{1}{n} \log s_n)$, respectively.
In particular, we define the \textit{type-I error rate limit} and the \textit{type-II error rate limit}
for a sequence $\Pi:=(\Pi_n)$ of quantum measurements (where, as mentioned, each orthogonal projection $\Pi_n$ corresponds to the
alternative hypothesis) as
\bea
\alpha_R(\Pi) &:=& \lim_{n\to\infty} \left(-\frac{1}{n}\log \alpha_n\right)
= \lim_{n\to\infty} \left(-\frac{1}{n}\log \trace[\rho^{\otimes n}\Pi_n]\right)\\
\beta_R(\Pi) &:=& \lim_{n\to\infty} \left( -\frac{1}{n}\log \beta_n\right)
= \lim_{n\to\infty} \left(-\frac{1}{n}\log \trace[\sigma^{\otimes n}(\id-\Pi_n)]\right),
\eea
if the limits exist. Otherwise we consider the limit inferior and the limit superior $\underline{\alpha}_R(\Pi)$
and $\overline{\alpha}_R(\Pi)$, respectively.
Similar definitions hold in the classical case.
\section{Bayesian Quantum Hypothesis Testing: Quantum Chernoff Bound}\label{sec:chernoff}
In this section we consider the Bayesian distinguishability problem. This means the goal is to minimise the average
error probability $P_e$, which is defined in (\ref{eq:pe}) and can be rewritten as
$P_e = \pi_1 - \trace[\Pi(\pi_1 \sigma - \pi_0 \rho)]$.
By the Neyman-Pearson Lemma, the optimal test is given by the projector $\Pi^*$ onto the range of
$(\pi_1 \sigma - \pi_0 \rho)_+$, and the obtained minimal error probability is given by
\beas
P_{e}^*
&=& \pi_1 - \trace[(\pi_1 \sigma - \pi_0 \rho)_+] \\
&=& \pi_1 - (\pi_1-\pi_0)/2 - \trace[|\pi_1 \sigma - \pi_0 \rho|/2] \\
&=& \frac{1}{2}\left(1 - \|\pi_1 \sigma - \pi_0 \rho\|_1 \right),
\eeas
where $\|A\|_1=\trace |A|$ is the trace norm.
We will call $\Pi^*$ the Holevo-Helstrom projector.

Next, note that the optimal test to discriminate $\rho$ and $\sigma$ in the case of $n$ copies enforces the use of
joint measurements. However, the particular permutational
symmetry of $n$-copy states guarantees that the optimal collective measurement can be implemented
efficiently (with a polynomial-size circuit) \cite{bacon04}, and hence that the minimum probability of error is
achievable with a reasonable amount of resources.

We need to consider the quantity
\be\label{eq:pendef}
P_{e,n}^*:=(1-\|\pi_1 \sigma^{\otimes n} - \pi_0 \rho^{\otimes n}\|_1)/2.
\ee
It turns out that  $P_{e,n}^*$ vanishes exponentially fast as $n$ tends to infinity.
The theorem below provides the asymptotic value of the exponent $-\frac{1}{n} \log P^*_{e,n}$, i.e.\ the rate limit of $P_{e,n}^*$,
which turns out to be given by the \textit{quantum Chernoff distance}. This is our main result.
\begin{theorem}\label{th:chernoff}
For any two states $\rho$ and $\sigma$ on a finite-dimensional Hilbert space,
occurring with prior probabilities $\pi_0$ and $\pi_1$, respectively,
the rate limit of $P_{e,n}^*$, as defined by (\ref{eq:pendef}), exists and is equal to the \textit{quantum Chernoff distance} $\xi_{QCB}$
\be\label{eq:szkbnd}
\lim_{n\rightarrow\infty} \left(-\frac{1}{n}\log P_{e,n}^*\right)
= \xi_{QCB} := -\log\left(\inf_{0\leq s\leq 1}\trace\left(\rho^{1-s}\sigma^{s}\right)\right).
\ee
\end{theorem}
Because the product of two positive operators always has positive spectrum,
 the quantity $\trace [\rho^{1-s}\sigma^{s}]$ is well
defined (in the mathematical sense) and guaranteed to be real and non-negative for every $0\le s\le 1$.
As should be, the expression for  $\xi_{QCB}$ reduces to the classical
Chernoff distance $\xi_{CB}$ defined by (\ref{eq:cbcl}) when $\rho$ and $\sigma$ commute.

%
\subsection{Proof of Theorem \ref{th:chernoff}: Optimality Part}\label{sec:chernoffopt}
In this Section, we will show that the best discrimination is specified by the quantum Chernoff distance; that is,   $\xi_{QCB}$ is an upper bound
on
$$
\limsup_{n\rightarrow\infty} \left( -\frac{1}{n}\log P_{e,n}\right)
$$
for any sequence of tests $(\Pi_n )$ and $P_{e,n}:=\pi_1 -\textrm{Tr }[\pi_1\sigma^{\otimes n} - \pi_0 \rho^{\otimes n}]$.

The proof, which first appeared in \cite{szkola}, is essentially based on relating the quantum to the classical case by using a special mapping
from a pair of $d \times d$ density matrices $(\rho,\sigma)$
to a pair of probability distributions $(p,q)$ on a set of cardinality $d^2$.


Let the spectral decompositions of $\rho$ and $\sigma$ be given by
$$
\rho=\sum_{i=1}^d \lambda_i \ket{x_i}\bra{x_i},\quad
\sigma=\sum_{j=1}^d \mu_j \ket{y_j}\bra{y_j},
$$
where $(\ket{x_i})$ and $(\ket{y_j})$ are two orthonormal bases of eigenvectors and $(\lambda_i)$ and $(\mu_j)$
are the corresponding sets of eigenvalues of $\rho$ and $\sigma$, respectively.
Then we map these density operators to the $d^2$-dimensional vectors
\be\label{eq:mapdef}
p_{i,j} = \lambda_i |\langle x_i|y_j\rangle|^2,\quad
q_{i,j} = \mu_j |\langle x_i|y_j\rangle|^2,
\ee
with $1\le i,j\le d$.
This  mapping preserves a number of important properties:
\begin{proposition}\label{prop:map}
With $p_{i,j}$ and $q_{i,j}$ as defined in (\ref{eq:mapdef}), and $s\in\R$,
\bea
\trace[\rho^{1-s}\sigma^s] &=& \sum_{i,j} p_{i,j}^{1-s} q_{i,j}^s \label{property1} \\
S(\rho\|\sigma) &=& H(p\|q).\label{property2}
\eea
\end{proposition}
Here, $S(\rho\|\sigma)$ is the quantum relative entropy defined as
\bea
S(\rho\|\sigma) &:=& \left\{\begin{array}{l}
\trace[\rho(\log\rho-\log\sigma)], \mbox{ if }\supp\rho\le\supp\sigma \\[1mm]
+\infty,\mbox{ otherwise,}
\end{array}\right. \label{def:relent}
\eea
where $\supp \rho$ denotes the support projection of an operator $\rho$, and
$H(p\|q)$ is the classical relative entropy,
or Kullback-Leibler distance,
\bea
H(p\|q) &:=& \left\{\begin{array}{l}
\sum_{i,j} p_{i,j}(\log p_{i,j}-\log q_{i,j}), \mbox{ if } p \ll q \\[1mm]
+\infty,\mbox{ otherwise.}
\end{array}\right. \label{eq:crelent}
\eea
\textit{Proof.}
The proof proceeds by direct calculation. For example:
\beas
\trace[\rho^{1-s}\sigma^s]
&=& \sum_{i,j} \lambda_i^{1-s} \mu_j^s |\langle x_i|y_j\rangle|^2 \\
&=& \sum_{i,j} \lambda_i^{1-s} \mu_j^s |\langle x_i|y_j\rangle|^{2(1-s)} |\langle x_i|y_j\rangle|^{2s} \\
&=& \sum_{i,j} p_{i,j}^{1-s} q_{i,j}^s.
\eeas
\qed

\bigskip

A direct consequence of identity (\ref{property1}) is that $p$ and $q$ are normalised if $\rho$ and $\sigma$ are.
Furthermore, tensor powers are preserved by the mapping; that is, if $\rho$ and $\sigma$ are mapped to
$p$ and $q$, then
$\rho^{\otimes n}$ is mapped to $p^{\otimes n}$
and $\sigma^{\otimes n}$ to $q^{\otimes n}$.

Now define the classical and quantum average (Bayesian) error probabilities $P_{e,c}$ and $P_{e,q}$ as
\bea
P_{e,c}(\phi,p,\pi_0,q,\pi_1)        &:=& \sum_i [\pi_0 \phi(i) p_i + \pi_1 (1-\phi(i)) q_i] \\
P_{e,q}(\Pi,\rho,\pi_0,\sigma,\pi_1) &:=& \trace[\pi_0 \Pi\rho + \pi_1 (\id-\Pi)\sigma],
\eea
where $p,q$ are probability distributions,
$\rho,\sigma$ are density matrices, and $\pi_0,\pi_1$ are the respective prior probabilities of the two hypotheses.
Furthermore, $\phi$ is a non-negative test function $0\le \phi\le 1$, and $\Pi$ is a positive semi-definite contraction, $0\le \Pi\le\id$,
so that $\{\id-\Pi,\Pi\}$ forms a POVM.

The main property of the mapping that allows to establish optimality of the quantum Chernoff distance
is presented in the following Proposition.
\begin{proposition}\label{prop:err}
For all orthogonal projectors $\Pi$
and all positive scalars $\eta_0,\eta_1$ (not necessarily adding up to 1),
and for $p$ and $q$ associated to $\rho$ and $\sigma$ by the mapping (\ref{eq:mapdef}),
$$
P_{e,q}(\Pi,\rho,\eta_0,\sigma,\eta_1) \ge \frac{1}{2} \inf_{\phi} P_{e,c}(\phi,p,\eta_0,q,\eta_1),
$$
where the infimum is taken over all test functions $0\le\phi\le1$.
\end{proposition}
Note that we have replaced the priors by general positive scalars; this will be useful later on, in proving the
optimality of the Hoeffding bound.

\medskip

\noindent\textit{Proof.}
Since $\Pi$ is a projector, one has $\Pi=\Pi\Pi=\sum_j \Pi\ket{y_j}\bra{y_j}\Pi$, where the second equality
is obtained by inserting a resolution of the identity
$\id=\sum_j \ket{y_j}\bra{y_j}$.
Likewise, $\id-\Pi$ is also a projector, and using another resolution of the identity,
$\id=\sum_i \ket{x_i}\bra{x_i}$, we similarly get
$\id-\Pi=\sum_i (\id-\Pi)\ket{x_i}\bra{x_i}(\id-\Pi)$.
This yields
\beas
\trace[\Pi\rho] &=& \sum_i \lambda_i \trace[\Pi\ket{x_i}\bra{x_i}] \\
&=& \sum_{i,j} \lambda_i \trace[\Pi\ket{y_j}\bra{y_j}\Pi \ket{x_i}\bra{x_i}] \\
&=& \sum_{i,j} \lambda_i |\bra{x_i}\Pi\ket{y_j}|^2,
\eeas
and, similarly,
\beas
\trace[(\id-\Pi)\sigma]
&=& \sum_{i,j} \mu_j |\bra{x_i}\id-\Pi\ket{y_j}|^2.
\eeas
Then the quantum error probability is given by
\beas
P_{e,q} &=& \eta_0 \trace[\Pi\rho] + \eta_1 \trace[(\id-\Pi)\sigma] \\
&=& \sum_{i,j} \eta_0 \lambda_i |\bra{x_i}\Pi\ket{y_j}|^2 + \eta_1 \mu_j |\bra{x_i}\id-\Pi\ket{y_j}|^2.
\eeas
The infimum of the classical error probability $P_{e,c}$ is obtained when the test function $\phi$
equals the indicator function $\phi=\chi_{\{\eta_1 q>\eta_0 p\}}$ (corresponding to the maximum likelihood decision rule);
hence, the value of this infimum is given
by
\beas
\inf_\phi P_{e,c} &=& \sum_{i,j} \min(\eta_0 p_{i,j},\eta_1 q_{i,j}) \\
&=& \sum_{i,j} \min(\eta_0 \lambda_i,\eta_1 \mu_j) |\langle x_i|y_j\rangle|^2.
\eeas

For a fixed choice of $i,j$, let $a$ be the $2\times2$ non-negative diagonal matrix
$$
a := \twomat{\eta_0\lambda_i}{0}{0}{\eta_1\mu_j},
$$
and let $b$ be the 2-vector
$$
b:=(\bra{x_i}\Pi\ket{y_j},\bra{x_i}\id-\Pi\ket{y_j}).
$$
The $i,j$-term in the sum for $P_{e,q}$ can then be written as the inner product
$\langle b|a|b\rangle$.
Similarly, the factor $|\langle x_i|y_j\rangle|^2$ occurring in the $i,j$-term in the sum for $P_{e,c}$
can then be written as $|b_1+b_2|^2$.

Now we note that $\langle b|b\rangle = \|b \|_2^2$, while $|b_1+b_2|^2\le \|b \|_1^2$.
For $d$-dimensional vectors, the inequality $\|b\|_2\ge \|b\|_1/\sqrt{d}$ holds; in our case, $d=2$.
Together with the inequality $a\ge \min(\eta_0\lambda_i,\eta_1\mu_j)\id_2$ this yields
\be
\langle b|a|b\rangle \ge \min(\eta_0\lambda_i,\eta_1\mu_j) \langle b|b\rangle
\ge \min(\eta_0\lambda_i,\eta_1\mu_j) \frac{1}{2} |b_1+b_2|^2.
\ee
Therefore, we obtain, for any $i,j$,
$$
\eta_0 \lambda_i |\bra{x_i}\Pi\ket{y_j}|^2 + \eta_1 \mu_j |\bra{x_i}\id-\Pi\ket{y_j}|^2
\ge \frac{1}{2} \min(\eta_0 \lambda_i,\eta_1 \mu_j) |\langle x_i|y_j\rangle|^2.
$$
As this holds for any $i,j$, it holds for the sum over $i,j$, so that
a lower bound for the quantum error probability is given by
$$
P_{e,q} \ge \frac{1}{2}\sum_{i,j} \min(\eta_0 p_{i,j},\eta_1 q_{i,j}) = \frac{1}{2}\inf_\phi P_{e,c},
$$
which proves the Proposition.
\qed

\bigskip

Using these properties of the mapping, the proof of optimality of the quantum Chernoff bound is easy.

\bigskip

\noindent\textit{Proof of optimality of the quantum Chernoff bound.}
Let hypotheses $H_0$ and $H_1$, with priors $\pi_0$ and $\pi_1$, correspond to the product states $\rho^{\otimes n}$
and $\sigma^{\otimes n}$. Using the mapping (\ref{eq:mapdef}), these states are mapped to the probability distributions
$p^{\otimes n}$ and $q^{\otimes n}$. By Proposition \ref{prop:err}, the quantum error probability
is bounded from below as
\be
P_{e,q}(\Pi_n,\rho^{\otimes n},\pi_0,\sigma^{\otimes n},\pi_1)
\ge \frac{1}{2}\inf_{\phi_n} P_{e,c}(\phi_n,p^{\otimes n},\pi_0,q^{\otimes n},\pi_1).\label{eq:qcl}
\ee
By the classical Chernoff bound, the rate limit of the right-hand side is given by
$$
-\log\inf_{0\le s\le 1}\sum_{i,j}p_{i,j}^{1-s} q_{i,j}^s
$$
(provided the priors $\pi_0,\pi_1$ are non-zero) and this is, therefore, an
upper bound on the rate limit of the optimal quantum error probability.
By Proposition \ref{prop:map} the latter
expression is equal to
$-\log\inf_{0\le s\le 1}\trace[\rho^{1-s}\sigma^s]$, which is what we set out to prove.
\qed

\bigskip

In a similar way one can prove the converse part of the quantum Hoeffding bound by relating it
to the classical problem in the sense of (\ref{eq:mapdef}), as already noted by Nagaoka in \cite{nagaoka06}.
This will be discussed in Section \ref{sec:hoeffdingopt}.

%
\subsection{Proof of Theorem \ref{th:chernoff}: Achievability Part}\label{sec:chernoffach}

In this Section, we prove the achievability of the quantum Chernoff bound, which is the statement
that the error rate limit $\lim_{n\rightarrow\infty} \left( -\frac{1}{n}\log P_{e,n}^*\right) $
is not only bounded above by, but is actually equal to the quantum Chernoff distance
$\xi_{QCB}$.
This can directly be inferred
from the following matrix inequality, which first made its appearance in \cite{spain}:
\begin{theorem}\label{th:1}
Let $a$ and $b$ be positive semi-definite operators, then for all $0\le s\le 1$,
\be\label{eq:main}
\trace[a^s b^{1-s}] \ge \trace[a+b-|a-b|]/2.
\ee
\end{theorem}
Note that inequality (\ref{eq:main}) is also interesting from a purely matrix analytic point of view, as it relates the
trace norm to a multiplicative quantity in a highly nontrivial and very useful way.

If we specialise this Theorem to states, $a=\sigma$ and $b=\rho$, with $\trace\rho=\trace\sigma=1$,
we obtain
$$
Q_s+T\ge1,\qquad 0\le s\le 1,
$$
where $Q_s:=Q_s(\rho,\sigma) := \trace[\rho^{1-s}\sigma^s]$ and
$T:=T(\rho,\sigma):=\|\rho-\sigma\|_1/2$ is the trace norm distance.

As an aside it is interesting to note that the inequality
$Q_s+T\ge1$ is \textit{strongly sharp}, which means that for any allowed value of $T$ one can find
$\rho$ and $\sigma$ that achieve equality.
Indeed, take the commuting density operators $\rho=\ket{0}\bra{0}$ and
$\sigma=(1-T)\ket{0}\bra{0}+T\ket{1}\bra{1}$, then
their trace norm distance is $T$, and $Q_s=1-T$.

\bigskip
\noindent\textit{Proof of achievability of the quantum Chernoff bound from Theorem \ref{th:1}}.

\noindent We will prove the inequality
\be\label{eq:converse}
\liminf_{n\rightarrow\infty}\left(-\frac{1}{n}\log P_{e,n}^*\right)\ge \xi_{QCB}.
\ee
Put $a=\pi_1\sigma^{\otimes n}$ and $b=\pi_0\rho^{\otimes n}$, so that the right-hand side of (\ref{eq:main})
turns into
$$
(1-\|\pi_1\sigma^{\otimes n}-\pi_0\rho^{\otimes n}\|_1)/2 = P_{e,n}^*.
$$
The logarithm of the left-hand side of inequality (\ref{eq:main}) simplifies to
$$
\log(\pi_0^{1-s}\pi_1^{s})+n\log\left(\trace[\rho^{1-s}\sigma^{s}]\right).
$$
Upon dividing by $n$ and taking the
limit $n\to\infty$, we obtain $\log Q_s$,
independently of the priors $\pi_0$, $\pi_1$
(as long as the priors are not degenerate, i.e.\ are different from 0 or 1). Then (\ref{eq:converse}) follows
from the fact that the inequality
\begin{eqnarray*}
\liminf_{n\rightarrow\infty}\left(-\frac{1}{n}\log P_{e,n}^*\right)\ge - \log Q_s
\end{eqnarray*}
holds for all $s \in [0,1]$ and we can replace the right-hand side by $\xi_{QCB}$.\qed

\bigskip

\noindent\textit{Proof of Theorem \ref{th:1}}.

\noindent 
The left-hand and right-hand sides of (\ref{eq:main}) look very disparate, but they can nevertheless
be brought closer together
by expressing $a+b-|a-b|$ in terms of the positive part $(a-b)_+$.
The inequality (\ref{eq:main}) is indeed equivalent to
\begin{eqnarray} \label{ineq-equivalent}
\trace[ a-a^s b^{1-s}] &\le& \trace [  a-(a+b-|a-b|)/2] \nonumber \\
&=& \trace[ (a-b+|a-b|)/2] \nonumber \\
&=& \trace[ (a-b)_+].
\end{eqnarray}

At this point we mention another equivalent formulation of this inequality, which will be used later in the proof of
the achievability of the quantum Hoeffding bound. With $\Pi$ the projector on the range of $(a-b)_+$, we can write:
\be\label{eq:hay}
\trace[a^s b^{1-s}] \ge \trace[ \Pi b+(\id-\Pi)a].
\ee

What we do next is strengthening the inequality (\ref{ineq-equivalent}) by replacing its left-hand side by an upper bound,
and its right-hand side by a lower bound.
Since, for any self-adjoint operator $H$, we have $H\le H_+$, we can write
\beas
\trace[a-a^s b^{1-s}] = \trace[a^s(a^{1-s}-b^{1-s})]
&\le& \trace[a^s(a^{1-s}-b^{1-s})_+] \\
&=& \trace[a^s \Pi^{(s)}(a^{1-s}-b^{1-s})] \\
&=& \trace[\Pi^{(s)}(a-b^{1-s}a^s)],
\eeas
where $\Pi^{(s)}$ is the projector on the range of $(a^{1-s}-b^{1-s})_+$.
Likewise,
$$
\trace [\Pi^{(s)}(a-b)]\le \trace[(a-b)_+],
$$
because $\trace [(a-b)_+]$ is the maximum of $\trace [\Pi(a-b)]$ over all orthogonal projections $\Pi$.
Inequality (\ref{eq:main}) would thus follow if, for that particular $\Pi^{(s)}$,
$$
\trace[\Pi^{(s)}(a-b^{1-s}a^s)] \le \trace \Pi^{(s)}(a-b).
$$
The benefit of this reduction is obvious, as
after simplification we get the much nicer statement
$$\trace[\Pi^{(s)} b^{1-s} (a^s-b^s)] \ge 0.$$
Equally obvious, though, is the risk of this strengthening; it could very well be a false statement.
Nevertheless, we show its correctness below.

It is interesting to note the meaning here of this strengthening in the context of the optimal hypothesis test, i.e.\
when $a=\sigma^{\otimes n}$ and $b=\rho^{\otimes n}$.
While the Holevo-Helstrom projectors $\Pi^*_n$ are optimal for every finite value of $n$, we can use other projectors
that are suboptimal but reach optimality in the asymptotic sense.
Here we are indeed using $\Pi^{(s^*)}$, the projector on the range of $(a^{1-s^*}-b^{1-s^*})_+$,
where $s^*$ is the minimiser of $\trace[\rho^{1-s}\sigma^s]$ over $[0,1]$, if it exists. Otherwise we have to use the Holevo-Helstrom projector.

In the next few steps we will further reduce the statement by reformulating the matrix powers in terms
of simpler expressions.
One can immediately absorb one of them into $a$ and $b$ via appropriate substitutions.
As we certainly don't want a power appearing in the definition of the projector $\Pi^{(s)}$, we are led to apply
the substitutions
$$A=a^{1-s},\quad B=b^{1-s},\quad t=s/(1-s).$$
This yields a value of $t$ between 0 and 1 only when $0\le s\le 1/2$. However, this is no restriction since
the case $1/2\le s\le 1$ can be treated in a completely similar way after applying an additional substitution $s\to 1-s$.

Inequality (\ref{eq:main}) is thus implied by the Lemma below, which ends the proof of Theorem \ref{th:1}.\qed

\begin{lemma}
For matrices $A,B\ge0$, a scalar $0\le t\le 1$,
and denoting by $P$ the projector on the range of $(A-B)_+$,
the following inequality holds:
\be\label{eq:lem4}
\trace[PB(A^t-B^t)]\ge0.
\ee
\end{lemma}
\textit{Proof.}
To deal with the $t$-th matrix power,
we use an integral representation (see, for example \cite{bhatia} (V.56)).
For scalars $a\ge0$ and $0\le t\le 1$,
$$
a^t = \frac{\sin(t\pi)}{\pi} \int_0^{+\infty} dx\,\,x^{t-1}\,\, \frac{a}{a+x}.
$$
For other values of $t$ this integral does not converge.
This integral can be extended to positive operators in the usual way:
$$
A^t = \frac{\sin(t\pi)}{\pi} \int_0^{+\infty} dx\,\, x^{t-1}\,\,A(A+x\id)^{-1}.
$$
To deal with non-invertible $A$ (arising when the states $\rho$ and $\sigma$ are not faithful),
we define $\lim_{x\to 0}A(A+x\id)^{-1}=\id$.

The potential benefit of this integral representation is that statements about the integral might follow
from statements about the integrand, which is a simpler quantity.

Applying the integral representation to $A^t$ and $B^t$, we get
$$
\trace[PB(A^t-B^t)] = \frac{\sin(t\pi)}{\pi} \int_0^{+\infty} dx\,\, x^{t-1}
 \trace[PB(A(A+x)^{-1}-B(B+x)^{-1})].
$$
If the integrand is positive for all $x>0$ (it is zero for $x=0$), then the whole integral is positive.
The Lemma follows if indeed we have
$$\trace[PB(A(A+x)^{-1}-B(B+x)^{-1})]\ge 0.$$

As a further reduction, we note that a difference can be expressed as an integral of a derivative:
$$
f(a)-f(b) = f(b+(a-b)) - f(b) = \int_0^1 dt\,\, \frac{d}{dt}f(b+(a-b)t).
$$
Here, we will apply this to the expression $A(A+x)^{-1}-B(B+x)^{-1}$.
Let $\Delta=A-B$. Then
\beas
A(A+x)^{-1}-B(B+x)^{-1} &=& \int_0^1 dt \,\,\frac{d}{dt}(B+t\Delta)(B+t\Delta+x)^{-1}.
\eeas
The potential benefit is again that the required statement might follow from a statement about the integrand,
which is a simpler quantity provided one is able to calculate the derivative explicitly.
In this case we are not dealing with a stronger statement, because the statement has to hold for the derivative anyway
(when $A$ is close to $B$).

In the present case, we can indeed calculate the derivative:
\beas
\frac{d}{dt}(B+t\Delta)(B+t\Delta+x)^{-1} &=& x\,\,(B+t\Delta+x)^{-1}\,\,\Delta\,\,(B+t\Delta+x)^{-1}.
\eeas
Therefore,
\beas
\lefteqn{\trace[PB(A(A+x)^{-1}-B(B+x)^{-1})]} \\
&=& x\,\int_0^1 dt \,\,\trace[PB(B+t\Delta+x)^{-1}\Delta(B+t\Delta+x)^{-1}].
\eeas
Again, if the integrand is positive for $0\le t\le 1$,
the whole integral is positive.
Absorbing $t$ in $\Delta$ we need to show, with $P$ the projector on $\Delta_+$:
$$
\trace[PB\,V\,\Delta\,V]\ge0,\qquad\mbox{where } V:=(B+\Delta+x)^{-1}\ge0.
$$

After all these reductions, the statement is now in sufficiently simple form to allow the final attack.
Since $B=V^{-1}-x-\Delta$, we have $BV\Delta V = \Delta(V-V\Delta V)-xV\Delta V$.
Positivity of $B$ implies $VBV = V-V\Delta V-xV^2\ge0$, thus $V-V\Delta V \ge x V^2$.
Furthermore, since $P\Delta = \Delta_+\ge0$,
\beas
\trace[PBV\Delta V]
&=& \trace[P(\Delta(V-V\Delta V)-xV\Delta V)] \\
&=& \trace[\Delta_+(V-V\Delta V)]-x\trace[PV\Delta V] \\
&\ge& x(\trace[\Delta_+ V^2]-\trace[PV\Delta V]).
\eeas
Because $\id\ge P\ge0$, $\Delta_+\ge0$, and $\Delta_+ \ge \Delta$,
$$
\trace[\Delta_+ V^2]=\trace[V \Delta_+ V] \ge \trace[P(V\Delta_+ V)] \ge \trace[P(V\Delta V)].
$$
The conclusion is that, indeed, $\trace[PBV\Delta V]\ge0$, which proves the Lemma.
\qed

\section{Properties of the Quantum Chernoff Distance\label{sec:properties}}
In this Section, we study the non-logarithmic variety $Q$ of the quantum Chernoff distance $\xi_{QCB}$, i.e.
\begin{eqnarray}\label{def:Q}
Q(\rho,\sigma):=\inf_{0\le s\le 1} \trace[\rho^{1-s} \sigma^{s}],
\end{eqnarray}
where $\rho, \sigma$ are density operators on a fixed finite-dimensional Hilbert space $\mathcal{H}$.
 All properties of $\xi_{QCB}=-\log Q$ can
readily be derived from $Q$.
It will turn out that $\xi_{QCB}$ is not a metric, since it violates the triangle inequality, but it  has a lot of properties required
of a distance measure on the set of density operators.

\subsection{Relation to Fidelity and Trace Distance}
The Uhlmann fidelity $F$ between two states is defined as
\be
F(\rho,\sigma) := \|\rho^{1/2} \sigma^{1/2}\|_1 = \trace [(\rho^{1/2}\sigma\rho^{1/2})^{1/2}].
\ee
Here, the latter formula is best known, but the first one is easier and makes the symmetry under interchanging
arguments readily apparent.
The Uhlmann fidelity can be regarded as the quantum generalisation of the so-called Hellinger affinity \cite{vanderVaart} defined as
$B(p_0,p_1):=\sum_i \sqrt{p_0(i) p_1(i)}$, where $p_0$ and $p_1$ are classical distributions.
It is an upper bound on $Q$, which can be shown as follows.
By definition, for any fixed value of $s \in [0,1]$, $Q_s=\trace[\rho^{1-s} \sigma^{s}]$ is an upper
bound on $Q$. In particular, this is true for $s=1/2$. Furthermore, by replacing the trace with the trace norm $\| \cdot \|_1$, we get an
even higher upper bound. Indeed,
\be
Q\le\trace[\rho^{1/2}\sigma^{1/2}]
= \|\rho^{1/4}\sigma^{1/2}\rho^{1/4}\|_1
\le \|\rho^{1/2}\sigma^{1/2}\|_1=F.\label{eq:QF}
\ee
In the last inequality we have used the fact (\cite{bhatia}, Prop.\ IX.1.1) that for any
unitarily invariant norm $|||AB|||\le |||BA|||$ if $AB$ is normal. In particular, consider
the trace norm, with $A=\rho^{1/4}\sigma^{1/2}$ and $B=\rho^{1/4}$.

For a pair of density operators the trace distance $T$ is defined by \[T(\rho,\sigma):= \frac{1}{2} \|\rho-\sigma \|_1.\]
Fuchs and van de Graaf \cite{fuchs} proved the following relation between $F$ and $T$:
\be\label{eq:TF}
(1-F)^2\le T^2\le 1-F^2.
\ee
Combining this with inequality (\ref{eq:QF}) yields the upper bound
\be\label{eq:QT}
Q^2+T^2\le 1.
\ee
Recall the relation $1-T \leq Q$, following from Theorem \ref{th:1}. Then combining everything yields the chain of inequalities
\be
1-\sqrt{1-F^2} \le 1-T\le Q\le F\le \sqrt{1-T^2}.
\ee
There is a sharper lower bound on $Q$ in terms of $F$, namely
\be\label{eq:QFL}
F^2\le Q.
\ee
This bound is strongly sharp, as it becomes an equality when one of the states is pure \cite{kargin}.
Indeed, for $\rho=|\psi\rangle\langle\psi|$, the minimum of the expression $\trace[\rho^{1-s}\sigma^{s}]$ is
obtained for $s=1$ and reduces to $\langle\psi|\sigma|\psi\rangle$, while $F$ is given by the square root of this expression.

We prove (\ref{eq:QFL}) in Appendix \ref{appa}, where we also give an alternative proof of the upper bound $Q\le \sqrt{1-T^2}$.
Both proofs go through in countably infinite dimensions.

\subsection{Range of $Q$}
The maximum value $Q$ can attain is 1, and this happens if and only if $\rho=\sigma$. This follows, for example, from
the upper bound $Q^2+T^2\le 1$. The minimal value is 0, and this is only attained for pairs of orthogonal states,
i.e.\ states such that $\trace \rho\sigma=0$. Consequently the range of the Chernoff distance is $[0, \infty]$
and the infinite value is attained on orthogonal states; this has to be contrasted with the relative
entropy, where infinite values are obtained whenever the states have a different support.

\subsection{Triangle inequality}
As already mentioned, on the set of pure states we have the identity $Q= F^2$. The Uhlmann fidelity $F$ does not obey the triangle inequality;
however it can be transformed into a metric by going over to $\arccos F$, while the Chernoff distance on pairs of pure states
is equal to $\xi_{QCB}=- \log Q= -2\log F$.

When considering the triangle inequality for $\xi_{QCB}$, one should note
first that in the classical case, the classical expression $\xi_{CB}$ should
be expected to behave like a \textit{squared} metric, similarly to the
relative entropy or Kullback-Leibler distance. Indeed consider two laws from
the normal shift family $N(\mu,1),$ $\mu\in\mathbb{R}$; then it is easy to see
that $\xi_{CB}=\left(  \mu_{1}-\mu_{2}\right)^{2}/8$. Thus $\xi_{CB}$
defines a squared metric on the normal shift family, which will not satisfy the
triangle inequality due to the square, but $\sqrt{\xi_{CB}}$ will. However
$\sqrt{\xi_{CB}}$ does not satisfy the triangle inequality in the general
case. To see this, let $Be(\varepsilon)$ be the Bernoulli law with parameter
$\varepsilon\in\lbrack0,1]$. Some computations show that $\xi_{CB}\left(
Be(1/2),Be(\varepsilon)\right)  \rightarrow\log2$ and $\xi_{CB}\left(
Be(\varepsilon),Be(1-\varepsilon)\right)  \rightarrow\infty$ as $\varepsilon
\rightarrow0$. As a consequence we have, for $\varepsilon$ small enough,
\[
\xi_{CB}^{1/2}\left(  Be(\varepsilon),Be(1-\varepsilon)\right)  >\xi
_{CB}^{1/2}\left(  Be(\varepsilon),Be(1/2)\right)  +\xi_{CB}^{1/2}\left(
Be(1/2),Be(1-\varepsilon)\right)
\]
contradicting the triangle inequality.

\subsection{Convexity of $Q_s$ as a function of $s$}
The target function $s\mapsto Q_s=\trace[\rho^{1-s} \sigma^{s}]$ in the variational formula defining $Q$ has the useful property to
be convex in $s\in[0,1]$  in the sense of Jensen's inequality:
$Q_{ts_1+(1-t)s_2}\le tQ_{s_1}+(1-t)Q_{s_2}$ for all $t \in [0,1]$. This
implies that a local minimum is automatically the global one, which is an important
benefit in actual calculations.

Indeed, the function $s\mapsto x^{1-s} y^{s}$ is analytic for positive scalars $x$ and $y$, and in this case its convexity may be
easily confirmed by calculating the second
derivative $x^{1-s} y^{s} (\log y-\log x)^2$, which is non-negative. If one of the parameters, say $x$, happens to be $0$,
then $s\mapsto x^{1-s} y^{s}$ is a constant function equal to $0$ for $s \in [0,1)$ and equal to $1$ at $s=1$. Hence, it is  still
convex, albeit discontinuous.
Consider then a basis with respect to which the matrix representation of $\rho$ is diagonal
$$\rho=\diag(\lambda_1,\lambda_2,\ldots).
$$
Let the matrix representation of $\sigma$ (in that basis) be given by
$$\sigma=U \diag(\mu_1,\mu_2,\ldots) U^*,
$$
where $U$ is a unitary matrix.
Then
$$
\trace[\rho^{1-s} \sigma^{s}] = \sum_{i,j}\lambda_i^{1-s} \mu_j^{s} |U_{ij}|^2.
$$
As this is a sum with positive weights of convex terms $\lambda_i^{1-s} \mu_j^{s}$, the sum itself is also convex in $s$.

\subsection{Joint concavity of $Q$ in $(\rho,\sigma)$}
By Lieb's theorem \cite{lieb},  $\trace[\rho^{1-s} \sigma^{s}]$ is
jointly concave on pairs of density operators $(\rho, \sigma)$ for each fixed $s \in \mathbb{R}$. Since $Q$ is the point-wise
minimum of $\trace[\rho^{1-s}\sigma^{s}]$ over $s\in[0,1]$, it is itself jointly concave as well.
Hence the related quantum Chernoff distance is jointly convex, just like the relative entropy.

\subsection{Monotonicity under CPT maps}
From the joint concavity one easily derives the following monotonicity
property: for any completely positive trace preserving (CPT) map $\Phi$ on the $C^*$-algebra $\mathcal{B}(\mathcal{H})$
of linear operators, one has
\be
Q(\Phi(\rho),\Phi(\sigma))\ge Q(\rho,\sigma).
\ee
To prove this, one first notes that $Q$ is invariant under unitary
conjugations, i.e.
$$
Q(U\rho U^*,U\sigma U^*) = Q(\rho,\sigma).
$$
Secondly, $Q$ is invariant under addition of an ancilla system: for any density operator $\tau$ on a finite-dimensional
ancillary Hilbert space we have the identity
$$
Q(\rho\otimes\tau,\sigma\otimes\tau) = Q(\rho,\sigma).
$$
This is because
$\trace[(\rho\otimes\tau)^{1-s}(\sigma\otimes\tau)^{s}] = \trace[\rho^{1-s} \sigma^{s}]\trace[\tau]$.
Exploiting the unitary representation of a CPT map, which is a special case of the Stinespring form,  the monotonicity
statement follows for general CPT maps if we can prove it for the partial trace map.
As noted by Uhlmann \cite{uhlmann,carlenlieb},
the partial trace map can be written as a convex combination of certain unitary conjugations.
Monotonicity of $Q$ under the partial trace then follows directly from its concavity and its unitary invariance.

\subsection{Continuity}
By the lower bound $Q+T\ge 1$, the distance measures $1-Q$ and $\xi_{QCB}$ are continuous in the sense that states that are
close in trace distance are also close w.r.t.\ $1-Q$ and w.r.t.\ $\xi_{QCB}$. Indeed, we have  $0\le 1-Q\le T$ and
$\xi_{QCB}=  -\log Q \leq - \log (1-T)=T + O(T^2)$.

\subsection{Relation of the Chernoff distance to the relative entropy}
In the classical case there is a striking relation between the Chernoff distance $\xi_{CB}$ and the relative entropy $H(\cdot \| \cdot)$.
It takes its simplest version if  the two involved  discrete probability distributions $p$ and $q$ have coinciding supports since then
$s \mapsto \log \sum_{x} p^{1-s}(x)q^s(x)=\log Q_s$ is analytic over $[0,1]$ and its infimum,  which  defines the Chernoff distance,
may be obtained simply by setting
\[0=(\log Q_s)'=H(p_{s} \| p)- H(p_{s}\| q)\]
(the prime denotes derivation w.r.t.\ $s$).
Here
\[p_s:=\frac{ p^{1-s}q^s}{\sum_{x} p^{1-s}(x)q^s(x)}\]
defines a parametric family of probability distributions interpolating between $p$ and $q$ as the
parameter $s$ varies between $0$ and $1$. In the literature, this family is called Hellinger arc.
It follows that the minimiser  $s^*\in [0,1]$ is uniquely determined by the identity
\begin{eqnarray}\label{midpoint-relent}
H(p_{s^*} \| q)= H(p_{s^*}\| p).
\end{eqnarray}
Furthermore, for any $s\in [0,1]$ we have:
\begin{eqnarray}\label{relent-Qs-p}
H(p_s \| p)= s (\log Q_s)'  - \log Q_s,
\end{eqnarray}
and similarly
\begin{eqnarray}\label{relent-Qs-q}
H(p_s \| q)= -(1-s) (\log Q_s)'  - \log Q_s,
\end{eqnarray}
This may be verified by direct calculation using
essentially the identity $ \log p^{1-s}q^s= \log p^{1-s}+\log q^{s}$. For the minimiser $s^*$ the formulas
(\ref{relent-Qs-p}) and (\ref{relent-Qs-q})  reduce to
\begin{eqnarray}\label{chernoff-relent}
H(p_{s^*}\| p)=H(p_{s^*}\| q)=\xi_{CB}(p,q).
\end{eqnarray}
In the generic case of possibly different supports of $p$ and $q$ one has to modify (\ref{midpoint-relent})
and (\ref{chernoff-relent}) slightly, see \cite{nussbaum}.

It turns out that in the quantum setting the minimiser $s^* \in [0,1]$ of $\inf_{s\in[0,1]}\log Q_s$ can be
characterised by a generalized version of (\ref{midpoint-relent}). However, the surely more remarkable
relation (\ref{chernoff-relent}) between the Chernoff distance $\xi_{CB}$ and the relative entropy seems to have no quantum counterpart.

We assume again that the involved density operators $\rho$ and $\sigma$ both have full support, i.e.\ are invertible.
Then $Q_s=\trace(\rho^{1-s}\sigma^{s})$ is an analytic function over $[0,1]$ and its local infimum over $[0,1]$,
which is a global minimum due to convexity, can be found by differentiating
$Q_s$ w.r.t.\ $s$:
\bea
\dds\trace[\rho^{1-s} \sigma^s] &=& -\trace[(\log\rho)\,\rho^{1-s}\,\sigma^s] + \trace[\rho^{1-s}\,\sigma^s\,\log\sigma] \nonumber\\
&=& -\trace[\rho^{1-s}\,\sigma^s\,\log\rho] + \trace[\rho^{1-s}\,\sigma^s\,\log\sigma].\label{eq:Qsdiff}
\eea
The infimum is therefore obtained for an $s \in [0,1]$ such that
\begin{eqnarray*}
\trace[\rho^{1-s}\sigma^{s}\log\rho] = \trace[\rho^{1-s}\sigma^{s}\log\sigma].
\end{eqnarray*}
This is equivalent to the condition
\begin{eqnarray}\label{eq:relent}
S(\rho_s||\rho)=S(\rho_s||\sigma),
\end{eqnarray}
where $S(\rho||\sigma)$ denotes the quantum relative entropy defined by (\ref{def:relent}) and $\rho_s$ is defined as
\begin{equation}
\rho_s=\frac{\rho^{1-s}\sigma^{s}}{\trace[\rho^{1-s}\sigma^{s}]}.\label{interpo}
\end{equation}
Note that $\rho_s$, with $s \in (0,1)$, is not a density operator, because it is not even self-adjoint
(except in the case of commuting $\rho$ and $\sigma$).
Nevertheless, as it is basically the product of two positive operators, it has positive spectrum,
and its entropy and the relative entropies used in (\ref{eq:relent}) are well-defined.
The value of $s$ for which both relative entropies coincide is the minimiser in the variational expression (\ref{def:Q}) for $Q$.

The family $\rho_{s}$, $s \in [0,1]$, can be considered as a quantum generalisation of the Hellinger arc
interpolating between the quantum states $\rho$ and $\sigma$,
albeit out of the state space, in contrast to the classical case.

When attempting to generalise relation (\ref{chernoff-relent}) to the quantum setting one has to  verify
(\ref{relent-Qs-p}) or (\ref{relent-Qs-q}) with density operators $\rho, \sigma$ replacing the probability distributions $p,q$.
This would require the identity $ \trace \rho_s \log \rho^{1-s}\sigma^{s} =\trace \rho_s (\log \rho^{1-s} + \log \sigma^{s})$
to be satisfied. However, this is not the case for arbitrary non-commutative density operators $\rho,\sigma$.
Thus the second identity in (\ref{chernoff-relent}) seems to be a classical special case only.

\section{Asymmetric Quantum Hypothesis Testing: Quantum Hoeffding Bound\label{sec:hoeffding}}
In this Section, we consider the applications of our techniques presented in
Section \ref{sec:chernoff} to the case of asymmetric quantum hypothesis testing.
More precisely, we consider a quantum generalisation of the Hoeffding bound and of Stein's Lemma.
\subsection{The Classical Hoeffding Bound}

The classical Hoeffding bound in information theory is due to Blahut
\cite{blahut} and Csisz\'{a}r and Longo \cite{csiszarLongo}. The corresponding
ideas in statistics were first put forward in the paper \cite{hoeffding} by
W.\ Hoeffding, from which the bound got its name. Some authors prefer the more
complete name of Hoeffding-Blahut-Csisz\'{a}r-Longo bound. In the following
paragraph we review the basic results in Blahut's terminology; at this point
we have to mention that many different notational conventions are in use
throughout the literature.

Let $p$ be the distribution associated with the null hypothesis, and $q$ the
one associated with the alternative hypothesis. \footnote{In \cite{blahut},
the null hypothesis corresponds to $H_{2}$, with distribution $q_{2}$,
\par
and the alternative hypothesis to $H_{1}$, with distribution $q_{1}$.}
Following \cite{blahut}, and for the purposes of this discussion, we initially
assume that $p$ and $q$ are equivalent (mutually absolutely continuous) on a
finite sample space. The Hoeffding bound gives the best exponential
convergence rate of the type-I error under the constraint that the rate limit
of the type-II error is bounded from below by a constant $r$, i.e.\ when the
type-II error tends to 0 sufficiently fast.

Blahut defines the \textit{error-exponent function} $e(r)$, $r\geq0$, with
respect to two probability densities $p$ and $q$ with coinciding supports, as
a minimisation over probability densities $x$:
\begin{eqnarray}\label{eq:er-def}
e(r)=\inf_{x}\{H(x\Vert p):H(x\Vert q)\leq r\},
\end{eqnarray}
where $H(\cdot\Vert\cdot)$ is again the classical relative entropy defined in
(\ref{eq:crelent}). This minimisation is a convex minimisation, since the
target function is convex in $x$, and the feasible set, defined by the
constraint $H(x\Vert q)\leq r$, is a convex set. Pictorially speaking, the
optimal $x$ is the point in the feasible set that is closest (as measured by
the relative entropy) to $p$. If $p$ itself is in the feasible set (i.e.\ if
$H(p\Vert q)\leq r$), then the optimal $x$ is $p$, and $e(r)=0$. Otherwise,
the optimal $x$ is on the boundary of the feasible set, in the sense that
$H(x\Vert q)=r$, and $e(r)>0$. Obviously, if $r=0$, the feasible set is the
singleton $\{q\}$, and $e(r)=H(q\Vert p)$.

\begin{figure}[ptb]
\includegraphics[width=12cm]{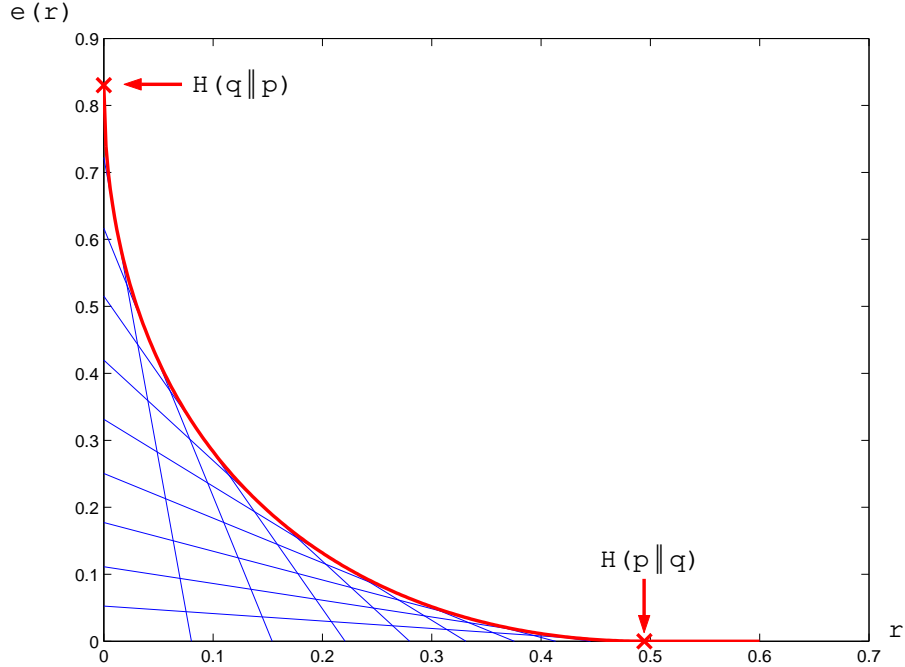}\caption{(Color online) Example
plot of the error-exponent function $e(r)$, eq.\ (\ref{eq:er-compx}), for the
distributions $p=(0.95,0.05)$ and $q=(0.5,0.5)$. The thick (red) line is the
graph of $e(r)$, while the thin (blue) lines are instances of the linear
function $(-rs-\log\sum_{k} q_{k}^{s} p_{k}^{1-s})/(1-s)$ for various values
of $s$, of which $e(r)$ is the point-wise maximum. For the chosen $p$ and $q$,
the value of $H(p\|q)=0.49463$ and the value of $H(q\|p)=0.83037$. }
\label{fig1}
\end{figure}

The error-exponent function is thus a non-increasing, convex function of
$r\geq0$, with the properties that $e(0)=H(q\Vert p)$ and $e(H(p\Vert q))=0$.
It can be expressed in a computationally more convenient format as%
\begin{equation}
e(r)=\sup_{0\leq s<\ 1}\frac{-rs-\log\sum_{k}q_{k}^{s}p_{k}^{1-s}}{1-s}
\label{eq:er-compx}
\end{equation}
An example is shown in Figure \ref{fig1}.

Let $\phi=(\phi_{n})$ be a sequence of test functions. Recall the notations
$\alpha_{R}(\phi)$ and $\beta_{R}(\phi)$ introduced in Section
\ref{sec:problem_formulation} for the rate limits (if they exist) of the
corresponding type-I and type-II errors, respectively:%
\[
\alpha_{R}(\phi)=\lim_{n\rightarrow\infty}-\frac{1}{n}\log\alpha_{n}%
(\phi),\;\beta_{R}(\phi)=\lim_{n\rightarrow\infty}-\frac{1}{n}\log\beta
_{n}(\phi)
\]
Then the classical HBCL Theorem can be stated as follows.

\begin{theorem}\label{thm:classHBCL}
(HBCL) Assume that $p,q$ are mutually absolutely continuous. Then for each
$r > 0$ there exists a sequence $\phi$ of test functions $  \phi
_{n}  $ such that the rate limits of the type-II and type-I errors
behave like $\beta_{R}(\phi)\geq r$ and $\alpha_{R}(\phi)=e(r)$. Moreover, for
any sequence $\phi$ such that $\alpha_{R}(\phi)$ and $\beta_{R}(\phi)$ both
exist, the relation $\beta_{R}(\phi)>r$ implies $\alpha_{R}(\phi)\leq e(r)$.
\end{theorem}

We remark that for sequences $\phi$ of test functions $\phi_{n}$ for which the
rate limits $\alpha_{R}(\phi)$ or $\beta_{R}(\phi)$ do not exist, the result
still applies to subsequences $(\phi_{n_{k}})$ along which both error rate
limits exist. The second part of the HBCL theorem is thus a statement about
all accumulation points of $\left(  -\frac{1}{n}\log\alpha_{n}(\phi),-\frac
{1}{n}\log\beta_{n}(\phi)\right)  $ for an arbitrary test sequence $\phi$.

Referring to Figure \ref{fig1}, the claim of this Theorem is that for any
sequence of test functions $\phi$ the point $(\beta_{R}(\phi),\alpha_{R}%
(\phi))$ cannot be above the graph of $e(r)$ over $r>0$ and for any point on
the graph over $r\geq0$ one can find a sequence $\phi$. Since $\beta_{R}%
(\phi)=0$ may correspond to the case where $\beta(\phi_{n})$ vanishes
subexponentially slowly as well as converges to a positive value, a rate limit
of type-I error $\alpha_{R}(\phi)$ larger than $e(0)=H(q\Vert p)$ is achievable.

The case $\beta_{R}(\phi)>$ $r\geq H(p\Vert q)$, where $e(r)=0$, can be shown
to correspond to $\alpha(\phi_{n})$ converging to $1$, rather than to $0$.
(This is basically the content of the so-called `Strong
Converse'.) In the case $\beta_{R}(\phi)=H(p\Vert q)$ a convergence of
$\alpha(\phi_{n})$ to $0$ is achievable, albeit only subexponantially slowly (this is due to Stein's Lemma.)

Note that in order to obtain a bound on $\beta_{R}$ under a constrained
$\alpha_{R}$ one just has to interchange $p$ and $q$ in the Theorem.
\subsection{Nonequivalent hypotheses}

The Chernoff and Hoeffding bounds have typically been treated in the
literature under a restrictive assumption that hypotheses $p,q$ are mutually
absolutely continuous (equivalent), cf., e.g., Blahut \cite{blahut}. As a
prerequisite for a quantum generalisation, unless one wants to limit oneself
to faithful states, one has to understand the classical Hoeffding bound for
nonequivalent hypotheses. For the Chernoff bound, a corresponding discussion
can be found in \cite{nussbaum} without restrictions on the underlying sample
space. Here we limit ourselves to finite sample spaces, thereby excluding
infinite relative entropies for equivalent measures $p,q$.

For probability measures $p,q$ on a finite sample space $\Omega$, let $D_{0}$
be the support of $p$, $D_{1}$ be the support of $q$ and $B=$ $D_{0}\cap
D_{1}$. Let $\psi_{0}=p\left(  B\right)  $, $\psi_{1}=q\left(  B\right)  $ and
note that $\psi_{0}>0,\psi_{1}>0$ unless the measures $p,q$ are orthogonal
(which we exclude for triviality). Define conditional measures given the set
$B$: $\tilde{p}\left(  \cdot\right)  =p\left(  \cdot|B\right)  $, $\tilde
{q}\left(  \cdot\right)  =q\left(  \cdot|B\right)  $. Note that $\tilde{p}$,
$\tilde{q}$ are equivalent measures; we may have $\tilde{p}=\tilde{q}$. We
consider hypothesis testing for a pair of product measures $p^{\otimes
n},q^{\otimes n}$.

Recall that a (nonrandomised) test is a mapping $\phi_{n}:\Omega^{n}%
\mapsto\{0,1\}.$ In our setting, only observations in either $D_{0}^{n}$ or
$D_{1}^{n}$ can occur, so we will modify the sample space to be $D_{0}^{n}%
\cup$ $D_{1}^{n}$. We will then establish relation of tests $\phi_{n}$ in the
original problem $p^{\otimes n}$ vs. $q^{\otimes n}$ to tests in the
`conditional' problem $\tilde{p}^{\otimes n}$ vs.\ $\tilde{q}^{\otimes n},$
i.e.\ to tests $\tilde{\phi}_{n}$ $:B^{n}\mapsto\{0,1\}$. Call a test $\phi
_{n}$ null admissible if it takes value $0$ on $D_{0}^{n}\setminus B^{n}$ and
value $1$ on $D_{1}^{n}\setminus B^{n}$. These tests correspond to the notion
that if a point in the sample space $\Omega^{n}$ is not in $B^{n}$, then it
identifies the hypothesis errorfree (either $p$ or $q$). We need only consider
null admissible tests; for any test there is a null admissible test with equal
or smaller error probabilities $\alpha_{n}$, $\beta_{n}$. The restriction $\phi_{n}|B^{n}$ gives a
test on $B^{n}$, i.e.\ in the conditional problem.

\begin{lemma}
\label{lem:null-admiss}There is a one-to-one correspondence between null
admissible tests $\phi_{n}$ in the original problem $p^{\otimes n}$ vs.
$q^{\otimes n}$ and tests $\tilde{\phi}_{n}$ in the conditional problem
$\tilde{p}^{\otimes n}$ vs. $\tilde{q}^{\otimes n}$, given by $\tilde{\phi
}_{n}=\phi_{n}|B^{n}$. The errror probabilities satisfy
\[
\alpha_{n}\left(  \phi\right)  =\psi_{0}^{n}\alpha_{n}\left(  \tilde{\phi
}\right), \beta_{n}\left(  \phi\right)  =\psi_{1}^{n}\beta_{n}\left(
\tilde{\phi}\right),
\]
where $\psi_0=p(B)$ and $\psi_1=q(B)$.
\end{lemma}
\begin{proof}
The first claim is obvious, if one takes into account that we took all tests
in the original problem to be mappings $\phi_{n}:D_{0}^{n}\cup$ $D_{1}%
^{n}\mapsto\{0,1\}$. For the relation of error probabilities, note that
$p^{\otimes n}(A)=$ $\psi_{0}^{n}\tilde{p}^{\otimes n}(A\cap B^{n})$,
$A\subset D_{0}^{n}\cup$ $D_{1}^{n}$ and therefore
\begin{align*}
\alpha_{n}\left(  \phi\right)   &  =\int\phi_{n}dp^{\otimes n}=\int_{B^{n}%
}\phi_{n}dp^{\otimes n}\text{ (by null admissibility)}\\
&  =\psi_{0}^{n}\int\phi_{n}d\tilde{p}^{\otimes n}=\psi_{0}^{n}\int_{B^{n}%
}\tilde{\phi}_{n}d\tilde{p}^{\otimes n}=\psi_{0}^{n}\alpha_{n}\left(
\tilde{\phi}\right)
\end{align*}
and analogously for $\beta_{n}\left(  \phi\right).$ \qed
\end{proof}

This result already allows to state the general Hoeffding bound in terms of
the error-exponent function for the conditional problem
\[
\tilde{e}(r)=\sup_{0\leq s<\ 1}\frac{-rs-\log\sum_{k}\tilde{q}_{k}^{s}%
\tilde{p}_{k}^{1-s}}{1-s}.
\]
Indeed, rate limits $\alpha_{R}(\phi)$ and $\beta_{R}(\phi)$ for a null
admissible test sequence $\phi$ exist if and only if they exist for the
corresponding test sequence $\tilde{\phi}$, and%
\begin{equation}
\alpha_{R}\left(  \phi\right)  =-\log\psi_{0}+\alpha_{R}\left(  \tilde{\phi
}\right)  \text{, }\beta_{R}\left(  \phi\right)  =-\log\psi_{1}+\beta
_{R}\left(  \tilde{\phi}\right)  .\label{eq:error-relat}%
\end{equation}

\begin{proposition}
\label{prop:reduc-to-tilde}Let $p,q$ be arbitrary probability measures on a
finite sample space. \newline(i) (achievability) For each $r\geq-\log\psi
_{1}$ there exists a sequence $\phi$ of test functions $\phi_{n}$ such that
the rate limits of the type-II and type-I errors behave like $\beta_{R}\left(
\phi\right)  \geq r$ and $\alpha_{R}\left(  \phi\right)  =-\log\psi_{0}%
+\tilde{e}(r+\log\psi_{1})$. For the case $0\leq r\leq-\log\psi_{1}$, there
is a sequence $\phi$ of test functions $\phi_{n}$ obeying  $-n^{-1}%
\log\beta_{n}\left(  \phi\right)  =-\log\psi_{1}$  and $\alpha_{n}\left(
\phi\right)  =0$ for every $n.$ \newline(ii) (optimality)  Consider  any
sequence $\phi$ such that $\alpha_{R}(\phi)$ and $\beta_{R}(\phi)$ both exist.
If $r\geq-\log\psi_{1}$ then  the relation $\beta_{R}(\phi)>r$ implies
$\alpha_{R}(\phi)\leq-\log\psi_{0}+\tilde{e}(r+\log\psi_{1})$.
\end{proposition}

Note that in (ii) the omission of the  case $0\leq r\leq-\log\psi_{1}$ means
that there is no upper bound on $\alpha_{R}(\phi),$ as shown by the
achievability part ($\alpha_{R}\left(  \phi\right)  $ has to be set equal to $\infty$
for a test of vanishing error probability $\alpha_{n}$).

\begin{proof}
(i) Assume $r\geq-\log\psi_{1}$ and take a test sequence $\tilde{\phi}_{n}$ in
the conditional problem $\tilde{p}^{\otimes n}$ vs.\ $\tilde{q}^{\otimes n}$
such that $\beta_{R}( \tilde{\phi})  \geq r+\log\psi_{1}$ and
$\alpha_{R}(\tilde{\phi})  =\tilde{e}(r+\log\psi
_{1})  $, which exists according to the HBCL theorem since  $\tilde
{p},\tilde{q}$ are mutually absolutely continuous. According to Lemma \ref{lem:null-admiss}, the corresponding null
admissible test $\phi_{n}$ satisfies
(\ref{eq:error-relat}) and hence $\beta_{R}\left(  \phi\right)  \geq r$ and
$\alpha_{R}\left(  \phi\right)  =-\log\psi_{0}+\tilde{e}(r+\log\psi_{1})$.
Furthermore, consider the test $\tilde{\phi}_{n}\equiv0$ in $\tilde{p}^{\otimes
n}$ vs.\ $\tilde{q}^{\otimes n}$. This has $\alpha_{n}(\tilde{\phi
}_n)  =0$ and  $\beta_{n}(\tilde{\phi}_n)  =1$, hence the
corresponding null admissible test $\phi_{n}$ has $\alpha_{n}(
\phi_n)  =0$ and $\beta_{n}(\phi_n)  =\psi_{1}^{n}$.

(ii) Using a reduction to the conditional problem $\tilde{p}^{\otimes n}$ vs.\
$\tilde{q}^{\otimes n}$ similar to the one above, the optimality part also
follows immediately from  the HBCL\ theorem. \qed
\end{proof}
\textit{Remark:} Consider the dual of the test used in the second part of
(i), i.e.\ the null admissible extension of the test $\tilde{\phi}_{n}\equiv1$.
This one obviously has  $\alpha_{n}\left(  \phi\right)  =\psi_{0}^{n}$ and
$\beta_{n}(\phi)  =0$. It can be used for achievability for
large $r$, i.e.\ it has $\beta_{R}(\phi)  =\infty$ and $\alpha
_{R}(\phi) =-\log\psi_{0}$.

It is possible to obtain a closed form expression for the Hoeffding bound,
using the error-exponent function defined for $r\geq0$ exactly  as in
(\ref{eq:er-compx}), for the case of nonequivalent $p,q$. The difference is
that we now have to admit a value $+\infty$ for certain arguments.

\begin{lemma}
\label{lem:repres}For general $p,q$, the error-exponent function $e(r)$
satisfies
\begin{eqnarray*}
e(r) =\left\{
\begin{array}{ll}
-\log \psi_0+ \tilde{e}(r+\log\psi_1), & \text{ for } r\leq -\log\psi_1 \\
\infty , & \text{ for } 0\leq r <-\log \psi_1.
\end{array}%
\right.
\end{eqnarray*}

\end{lemma}
\textit{Remark:} For two distinct $p,q$ it is  possible that $\tilde{p}
=\tilde{q}$. In that case $\tilde{e}(r)=0$ for $r\geq 0$. It follows that $e(r)=\infty$
for $r < -log \psi_1$ and $e(r)=- \log \psi_0$ for $r \geq -\log \psi_1$. This case will be
relevant in the quantum setting when the hypotheses will be represented by two non-orthogonal pure quantum states.
\begin{proof}
Assume $r\geq-\log\psi_{1}$ and set
\[
e_{s}(r)=\frac{-rs-\log Q_{s}}{1-s}%
\]
where $Q_{s}=\sum_{k}p_{k}^{1-s}q_{k}^{s}$. Let $\tilde{Q}_{s}=\sum_{k}%
\tilde{p}_{k}^{1-s}\tilde{q}_{k}^{s}$ and note $Q_{s}=\psi_{0}^{1-s}\psi
_{1}^{s}\tilde{Q}_{s}$. Hence
\begin{align*}
e_{s}(r)  & =\frac{-rs-(1-s)\log\psi_{0}-s\log\psi_{1}-\log \tilde{Q}_{s}}{1-s}\\
& =-\log\psi_{0}+\frac{-(r+\log\psi_{1})s-\log \tilde{Q}_{s}}{1-s}=-\log\psi
_{0}+\tilde{e}_{s}(r+\log\psi_{1})
\end{align*}
where $\tilde{e}_{s}$ is the analogue of the function $e_{s}(r)$ with $Q_{s}$
replaced by $\tilde{Q}_{s}$. Since $e(r)=\sup_{0\leq s<\ 1}$ $e_{s}(r)$ and
the analogue is true for $\tilde{e}_{s}$ and $\tilde{e}$, the claim follows in
the case $r\geq-\log\psi_{1}$.

Assume now  $0\leq r<-\log\psi_{1}$ and $\psi_{1}<1$, i.e.\  $-\log\psi_{1}>0$.
Clearly  we have   $Q_{s}\rightarrow\psi_{1}$ as $s\nearrow1$, hence
$-rs-\log Q_{s}\rightarrow-r-\log\psi_{1}>0$ as $s\nearrow1$. Hence
$\lim_{s\nearrow1}e_{s}(r)=\infty$, and since $e(r)=\sup_{0\leq s<\ 1}$
$e_{s}(r)$, we also have $e(r)=\infty$. \qed
\end{proof}

In conjunction with Proposition \ref{prop:reduc-to-tilde} we obtain a closed
form description of the Hoeffding bound for possibly nonequivalent measures
$p,q$, in terms of the original error-exponent function $e(r)$.

\begin{theorem}\label{th:hbcl-class-general}
Let $p,q$ be arbitrary probability measures on a finite sample space. Then the
statement of the  HBCL Theorem (Theorem \ref{thm:classHBCL}) is true,
where the error-exponent function defined in (\ref{eq:er-compx}) obeys
$e(r)=\infty$ for $0\leq r<-\log\psi_{1}$ if $\psi_{1}<1$.
\end{theorem}
We noted already that for  $e(r)=\infty$, the bound on $\alpha_{R}(\phi)$ is
achievable in the sense that a test exists having exactly $\alpha_{n}(\phi)=0$
for all $n$.

Using the properties of the rate function $\tilde{e}$ pertaining to equivalent
measures $\tilde{p},\tilde{q}$, as illustrated in Figure \ref{fig1}, and the
representation of Lemma \ref{lem:repres} we obtain the following description
of the general rate exponent function. In the interval $[0,-\log\psi_{1})$ it
is infinity. At $r=$ $-\log\psi_{1}$ it takes value  $e(r)=-\log\psi
_{0}+H(\tilde{q}\Vert\tilde{p})=H(\tilde{q}\Vert p)$. For $r \geq - \log \psi_1$
it is convex and non-increasing. More precisely, over the interval
$[-\log\psi_{1},-\log\psi_{1}+H(\tilde{p}\Vert\tilde{q})=H(\tilde{p}\Vert
q)]$ $e(r)$ is convex (even strictly convex) and monotone decreasing. Over the interval $[H(\tilde{p}\Vert
q),\infty)$ it is constant with value  $-\log\psi_{0}$. A  visual impression can be
obtained by imagining  the origin in Figure \ref{fig1} shifted to the point $\left(
-\log\psi_{1},-\log\psi_{0}\right)  $. This  picture will explicitly appear
in Figure \ref{fig2} below, in a situation further generalized  to two quantum
states with different supports.
\subsection{Quantum Hoeffding Bound}
In the quantum setting the error-exponent function $e(r)$ has to be replaced by a
function $e_Q:\R^+_0 \longrightarrow [0,\infty]$ given by
\be\label{eq:erq-defx}
e_Q(r) := \sup_{0\le s < 1} \frac{-rs-\log\trace \sigma^s \rho^{1-s}}{1-s}.
\ee
In view of Proposition \ref{prop:map}, $e_{Q}(r)$ coincides with the error-exponent
function $e(r)$ for the pair of probability distributions $(p,q)$
associated with $(\rho,\sigma)$ via relation (\ref{eq:mapdef}). Therefore, we
can use Lemma \ref{lem:repres} to describe properties of the function $e_{Q}(r)$,
or the remarks after Theorem \ref{th:hbcl-class-general}.

Recall that for a pair $(p,q)$, we defined a related pair of probability
distributions $(\tilde{p},\tilde{q})$ by conditioning $p$ and $q$, respectively, on the
intersection $B=D_{0}\cap D_{1}$ of the two support sets $D_{0}$ and $D_{1}$,
and also $\psi_{0}=p(B),$ $\psi_{1}=q(B)$. In the present context, in
accordance with (\ref{eq:mapdef}) we have%
\[
D_{0}=\left\{  (i,j):1\leq i,j\leq d,\;\lambda_{i}>0\right\}  ,\;D_{1}%
=\left\{  (i,j):1\leq i,j\leq d,\;\mu_{j}>0\right\}  .
\]
Let, as before, $\tilde{e}(r)$ be the error-exponent function pertaining to the
pair $(\tilde{p},\tilde{q})$ according to (\ref{eq:er-compx}). Then the quantum
error-exponent function $e_{Q}(r)$ for the hypotheses $\rho,\sigma$
may be represented simply by
\begin{eqnarray}\label{eQ=e}
e_{Q}(r) = e(r)=\left\{
\begin{array}{ll}
-\log \psi_0+ \tilde{e}(r+\log\psi_1), & \text{for } r\leq -\log\psi_1 \\
\infty , & \text{for } 0\leq r <-\log \psi_1.
\end{array}%
\right.
\end{eqnarray}
It obtains its characteristic properties from the classical function being convex
and monotone decreasing in the interval $\left[
-\log\psi_{1},H(\tilde{p}\Vert q)\right]  $ with $e(-\log\psi_{1})=H(\tilde
{q}\Vert p)$, and constant with value $-\log\psi_{0}$ in the interval
$[H(\tilde{p}\Vert q),\infty)$.
\begin{lemma}
\label{lem:pseudo-entropies}Let \textrm{supp }$\rho$, \textrm{supp }$\sigma$
be the support projections associated with $\rho,\sigma$. Then the critical points
and extremal values of $e_Q(r)$ may be expressed in a more direct way in terms of the density operators:
\begin{eqnarray*}
\psi_{0}=\mathrm{Tr }\left[\rho\;\mathrm{supp\ }\sigma\right],\quad \psi_1=\mathrm{Tr }\left[\sigma\;\mathrm{supp\ }\rho\right]
\end{eqnarray*}
and
\begin{eqnarray*}
H(\tilde{p}\Vert q)=S_{\sigma}(\rho\Vert\sigma)\quad H(\tilde{q}\Vert
p)=S_{\rho}(\sigma\Vert\rho),
\end{eqnarray*}
where the entropy type quantities on the right-hand side are defined as
\begin{align*}
S_{\sigma}(\rho\Vert\sigma)  :=\mathrm{Tr}\left[  \frac{\rho}{\psi_{0}%
}\;\left(  \log\frac{\rho}{\psi_{0}}-\log\sigma\right)  \mathrm{supp\ }%
\sigma\right],\\
S_{\rho}(\sigma\Vert\rho) :=\mathrm{Tr}\left[  \frac{\sigma}{\psi_{1}
}\;\left(  \log\frac{\sigma}{\psi_{1}}-\log\rho\right)  \mathrm{supp\ }
\rho\right].
\end{align*}
\end{lemma}
\begin{proof}
Note that for $B=D_{0}\cap D_{1}$ we have
\begin{align*}
\psi_{0}  &  =\sum_{(i,j)\in B}\lambda_{i}\left\vert \left\langle x_{i}%
|y_{j}\right\rangle \right\vert ^{2}=\sum_{i,j}\lambda_{i}\;\mathrm{sgn}%
(\mu_{j})\;\left\vert \left\langle x_{i}|y_{j}\right\rangle \right\vert ^{2}\\
&  =\sum_{i,j}\lambda_{i}\;\left\vert \left\langle x_{i}|\mathrm{sgn}(\mu
_{j})y_{j}\right\rangle \right\vert ^{2}=\sum_{i,j}\lambda_{i}\;\left\vert
\left\langle x_{i}|\left(  \mathrm{supp\ }\sigma\right)  y_{j}\right\rangle
\right\vert ^{2}\\
&  =\sum_{i,j}\lambda_{i}\;\left\vert \left\langle \left(  \mathrm{supp\ }%
\sigma\right)  x_{i}|y_{j}\right\rangle \right\vert ^{2}=\sum_{i}\lambda
_{i}\;\left\Vert \left(  \mathrm{supp\ }\sigma\right)  x_{i}\right\Vert
^{2}=\\
&  =\mathrm{Tr}\left[  \sum_{i}\lambda_{i}\left\vert \left(  \mathrm{supp\ }%
\sigma\right)  x_{i}\right\rangle \left\langle \left(  \mathrm{supp\ }%
\sigma\right)  x_{i}\right\vert \right]  =\mathrm{Tr}\left[  \rho
\;\mathrm{supp\ }\sigma\right]
\end{align*}
and analogously for $\psi_{1}$. Furthermore
\begin{align*}
H(\tilde{p}\Vert q)  &  =\sum_{(i,j)\in B}\tilde{p}_{i,j}\log\frac{\tilde
{p}_{i,j}}{q_{i,j}}=\sum_{(i,j)\in B}\lambda_{i}\left\vert \left\langle
x_{i}|y_{j}\right\rangle \right\vert ^{2}\frac{1}{\psi_{0}}\log\frac
{\lambda_{i}}{\mu_{j}\psi_{0}}\\
&  =\sum_{i,j} \mathrm{sgn}(\mu_{j}) \left\vert \left\langle x_{i}|y_{j}\right\rangle
\right\vert ^{2} \frac{\lambda_{i}}{\psi_{0}}\log\frac{\lambda_{i}}{\psi_{0}
}-\sum_{i,j}\mathrm{sgn}(\mu_{j}) \left\vert \left\langle x_{i}|y_{j}\right\rangle
\right\vert ^{2}\frac{\lambda_{i}}{\psi_{0}}\log\mu_{j}\\
&  =\mathrm{Tr}\left[  \frac{\sigma}{\psi_{1}}\;\left(  \log\frac{\sigma}%
{\psi_{1}}\right)  \mathrm{supp\ }\rho\right]  -\mathrm{Tr}\left[
\frac{\sigma}{\psi_{1}}\;\left(  \log\rho\right)  \mathrm{supp\ }\rho\right]
\end{align*}
where the third equality is analogous to the calculation in the proof of
Proposition \ref{prop:map}. \qed
\end{proof}
To shed some light on the entropy type quantity $S_{\sigma}(\rho \| \sigma)$,
note that it may be rewritten as a difference of usual (Umegaki's) relative entropies:
\begin{eqnarray*}\label{rel:pseudo-entropy_rel-entropy}
S_{\sigma}(\rho \| \sigma)= S(\frac{\rho}{\psi_0} \textrm{supp } \sigma \| \sigma)
- S(\frac{\rho}{\psi_0} \textrm{supp } \sigma \| \frac{\rho}{\psi_0}).
\end{eqnarray*}
This may be verified by direct calculations similar to those in the proof of Lemma \ref{lem:pseudo-entropies}.

The linear operator $\frac{\rho}{\psi_0} \textrm{supp } \sigma$ is
a kind of conditional expectation of $\rho$. While it is not self-adjoint,
the relative entropies on the right-hand side are well defined (in a mathematical sense) and real:
first, the entropy of $\frac{\rho}{\psi_0} \textrm{supp } \sigma$ is defined in terms of its spectrum,
which is positive and normalised to 1, hence giving a real, positive entropy, and second,
$\trace[\rho \,\,\textrm{supp } \sigma \log(\rho)]$ can be written as
$\trace[\textrm{supp } \sigma \rho\log\rho \,\,\textrm{supp } \sigma]$, from which it is evident that this term is also real.

It is easily seen from the above formula that $S_{\sigma}(\rho \| \sigma)$
coincides with $S(\rho \|\sigma )$ if $\sigma$ is a faithful
state, or more generally if $\mathrm{supp\ }\rho\leq\mathrm{supp\ }\sigma$.
Otherwise $S(\rho \| \sigma)=\infty$, while $S_{\sigma}(\rho\| \sigma)$ is finite.

Note also that $S_{\rho}(\sigma\Vert\rho)\geq-\log\psi_{0}$ and equality holds
if and only if it holds in $S_{\sigma}(\rho\Vert\sigma)\geq-\log\psi_{1}$.
This immediately follows from $S_{\rho}(\sigma\Vert\rho)+\log\psi_{0}=$
$H(\tilde{p}\Vert\tilde{q})$, which is seen from Lemma
\ref{lem:pseudo-entropies}. This happens in particular if both $\rho$ and
$\sigma$ are pure states. In this case there is only one pair $(i,j)$ where
both $\lambda_{i}>0$ and $\mu_{j}>0$, hence the set $B$ consists of one
element only. In this case we must have $\tilde{p}=\tilde{q},$ hence
$H(\tilde{p}\Vert\tilde{q})=H(\tilde{q}\Vert\tilde{p})=0$.

\begin{figure}[ptb]
\includegraphics[width=12cm]{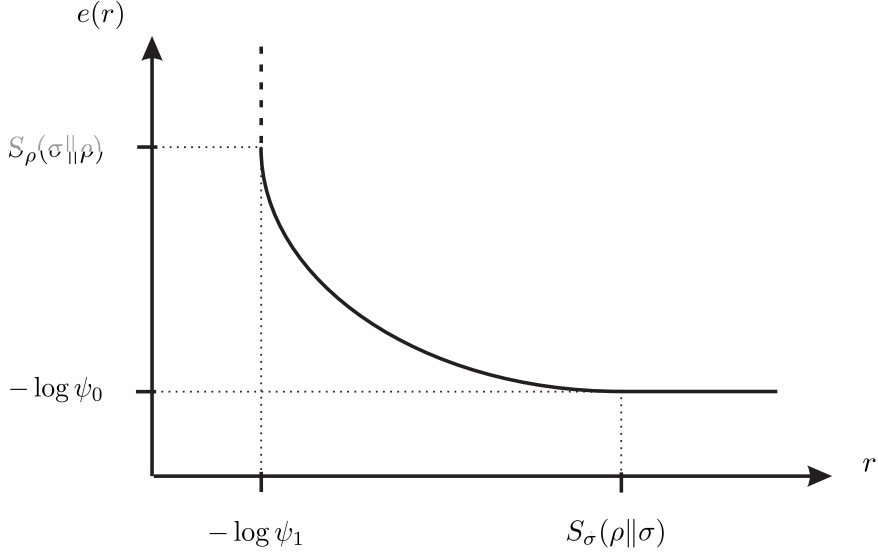}
\caption{Example plot of the quantum error-exponent function $e_Q(r)$ in the general case. }
\label{fig2}
\end{figure}

The general shape of the quantum error-exponent function $e_{Q}(r)$ is
represented in Figure \ref{fig2}.
If both $\rho$ and $\sigma$ are pure states then the
shape degenerates to `rectangular' form ($e_{Q}(r)=\infty$ or $e_{Q}(r)=-\log
\psi_{1}$).

A quantum generalisation of the HBCL Theorem then reads as follows.

\begin{theorem}
(Quantum HBCL)\label{th:qhoeff}For each $r>0$ there exists a sequence $\Pi$
of test projections $\Pi_{n}$ on $\mathcal{H}^{\otimes n}$ for
which the rate limits of type-I and type-II errors behave like $\alpha_{R}%
(\Pi)=e_{Q}(r)$ and $\beta_{R}(\Pi)\geq r$, respectively. Moreover, for any sequence $\Pi$
such that $\alpha_{R}(\Pi)$ and $\beta_{R}(\Pi)$ both exist, the relation
$\beta_{R}(\Pi)>r$ implies $\alpha_{R}(\Pi)\leq e_{Q}(r)$.
\end{theorem}

The statement of the quantum HBCL Theorem is that for every sequence $\Pi$
(for which both error rate limits exist) the point $(\beta_{R}(\Pi),\alpha
_{R}(\Pi))$ lies on or below the curve $e_{Q}(r)$ over $(0,\infty]$, and for
every point on the curve over the closed interval $[0,\infty]$ there is a
sequence $\Pi$ achieving it.

We remark that, just like (\ref{chernoff-relent}), the relationship
(\ref{eq:er-def}) seems to have no general quantum counterpart, even when both
states are faithful. In other words, there is no known subset of linear
operators $\tau$ with positive spectrum such that $e_{Q}(r)=\inf_{\tau
}\{S(\tau\Vert\rho):S(\tau\Vert\sigma)\leq r\}$.

To prove the quantum Hoeffding bound, the following lemmas are needed.
\begin{lemma}\label{lem:log}
For scalars $x,y>0$, bounds on $\log(x+y)$ are given by
\be
\max(\log x,\log y) \le \log(x+y) \le \max(\log x,\log y)+\log 2.
\ee
\end{lemma}
\textit{Proof.}
For the first inequality, put $x=e^a$ and $y=e^b$, and note
\beas
\log(e^a+e^b)&=&a+\log(1+e^{b-a}) \\
&\ge& a+\max(0,b-a) \\
&=&\max(a,b).
\eeas
The second inequality follows directly from the fact that the logarithm increases mono\-tonically,
so that $\log((x+y)/2) \le \log\max(x,y)$.
\qed

A direct consequence of this Lemma is
\begin{lemma}\label{lem:logseq}
For two scalar sequences $x_n,y_n>0$ with rate limits $x_R$ and $y_R$,
the rate limit of $x_n+y_n$ is given by
\be
\lim_{n\to\infty} -\frac{1}{n}\log(x_n+y_n) = \min(x_R,y_R).
\ee
\end{lemma}
\subsection{Proof of Optimality of the Quantum Hoeffding Bound\label{sec:hoeffdingopt}}
Again we use the mapping from the pair $(\rho,\sigma)$ to the pair $(p,q)$, so
that, by Proposition \ref{prop:map}, $e(r)=e_{Q}(r)$. From Proposition
\ref{prop:err} we have that for any sequence $\Pi$ of orthogonal projections
$\Pi_{n}$ and for any real value of the scalar $x$, for all
$n\in\mathbb{N}$ one as
\[
\alpha(\Pi_{n})+e^{-nx}\beta(\Pi_{n})\geq\frac{1}{2}\left(  \alpha(\phi
_{n})+e^{-nx}\beta(\phi_{n})\right),
\]
where $\phi_{n}$ are classical test functions corresponding to the maximum
likelihood decision rule, cf.\ the proof of Proposition \ref{prop:err}. Recall
that the type-I and type-II errors are defined as $\alpha(\phi_{n})=\sum
_{i}p_{i}^{n}\phi_{n}(i)$ and $\beta(\phi_{n})=\sum_{i}q_{i}^{n}(1-\phi
_{n}(i)).$

On taking the rate limit on the left side,  this gives
\[
\lim_{n\rightarrow\infty}-\frac{1}{n}\log\left(  \alpha(\Pi_{n})+e^{-nx}%
\beta(\Pi_{n})\right)  \leq\liminf_{n\rightarrow\infty}-\frac{1}{n}\log\left(
\alpha(\phi_{n})+e^{-nx}\beta(\phi_{n})\right)
\]
By possibly taking a subsequence, we can ensure that the rate limits
$\alpha_{R}(\phi)$, $\beta(\phi_{n})$ also exist. By Lemma \ref{lem:logseq}, the above  simplifies to
\begin{equation}
\min(\alpha_{R}(\Pi),x+\beta_{R}(\Pi))\leq\min(\alpha_{R}(\phi),x+\beta
_{R}(\phi)).\label{eq:up-bound}%
\end{equation}
Assume now that $\beta_{R}(\phi)\leq-\log\psi_{1}$. Then, by selecting $x<0$
and $\left\vert x\right\vert $ sufficiently large, we obtain $x+\beta_{R}%
(\Pi)\leq x+\beta_{R}(\phi)$ and hence $\beta_{R}(\Pi)\leq-\log\psi_{1}$.
Since $e_{Q}(r)=\infty$ for $r<\beta_{R}(\Pi)\leq-\log\psi_{1}$ according to the discussion above
Lemma \ref{lem:pseudo-entropies}, the claim $\alpha_{R}(\Pi)\leq
e_{Q}(r)$ holds trivially. Henceforth we assume that $\beta_{R}(\phi
)>-\log\psi_{1}$.

From the classical HBCL Theorem (more precisely, from Theorem
\ref{th:hbcl-class-general}), the right-hand side of (\ref{eq:up-bound}) is bounded above by
$\min(e(r),x+\beta_{R}(\phi))$,  for any $r$ with $-\log\psi_{1}\leq
r<\beta_{R}(\phi)$. Note that $e(r)$ is continuous for $r\geq-\log\psi_{1}$
(since it is monotonely nonincreasing and convex). By letting $r\nearrow
\beta_{R}(\phi)$ we obtain an upper bound $\min(e(r),x+r)$ with $r\geq
-\log\psi_{1}$.

We can now prove the optimality part of the quantum HBCL Theorem, using only
this upper bound plus the fact that $e(r)$ is monotonously decreasing.

The upper bound $\min(e(r),x+r)$ holds for some particular value $r$.   We
will find a further upper bound by maximizing over $r\geq-\log\psi_{1}$. For
this we have to distinguish two cases, depending on the value of $x.$

a)  At $r=-\log\psi_{1}$ we have $e(r)>x+r$. Since $e(r)$ is decreasing in $r$
and continuous, and $x+r$ is increasing, the maximum of $\min(e(r),x+r)$ is
obtained when $e(r)=x+r$. Let  $r^{\ast}(x)>-\log\psi_{1}$ be the  solution of
$x+r=e(r)$.  We now have that for any sequence of quantum measurements $\Pi$
and for any real value of the scalar $x$,
\[
\min(\alpha_{R}(\Pi),x+\beta_{R}(\Pi))\leq x+r^{\ast}(x)=e(r^{\ast}(x)).
\]

b) At $r=-\log\psi_{1}$ we have $e(r)\leq x+r$. Again by the properties of
$e(r)$ and $x+r$, the maximum of $\min(e(r),x+r)$ is $e(r^{\ast})$ is attained
for $r^{\ast}(x)=-\log\psi_{1}$. We then obtain the upper bound
\[
\min(\alpha_{R}(\Pi),x+\beta_{R}(\Pi))\leq e(r^{\ast}(x)).
\]
Now set $x=\alpha_{R}(\Pi)-\beta_{R}(\Pi)$, then both inequalities above yield
$\alpha_{R}(\Pi)\leq e(r^{\ast})$. Assume $r<\beta_{R}(\Pi)$; we intend to
show that this implies $\alpha_{R}(\Pi)\leq e(r)$.  Indeed, in both cases a)
and b) $r^{\ast}$ is such that
\[
e(r^{\ast})\leq x+r^{\ast}=\alpha_{R}(\Pi)-\beta_{R}(\Pi)+r^{\ast}<\alpha
_{R}(\Pi)-r+r^{\ast}%
\]
hence  $r^{\ast}-r\geq e(r^{\ast})-\alpha_{R}(\Pi)\geq0$. Therefore, from the
monotonicity of the error-exponent function follows $e(r^{\ast})\leq e(r)$
and we finally obtain $\alpha_{R}(\Pi)\leq e(r)=e_{Q}(r)$. \qed
\subsection{Proof of Achievability of the Quantum Hoeffding Bound\label{sec:hoeffdingach}}
The proof of achievability is mainly due to Hayashi \cite{hayashi06},
who used inequality (\ref{eq:hay}), which is obtained as
a byproduct of the proof of Theorem \ref{th:1}. However, we modify it
avoiding any implicit assumption that the involved quantum states are faithful; hence we prove
Theorem \ref{th:qhoeff} in full generality, which includes for example the case of two non-orthogonal pure states.

Let us fix an arbitrary $s \in (0,1)$, and set
\bea
a&=&e^{-nx}\sigma^{\otimes n} \label{eq:haya}\\
b&=&\rho^{\otimes n}, \label{eq:hayb}
\eea
where the value of $x$ will be chosen in due course.
Consider the sequence of POVMs $\{(\id-\Pi_n,\Pi_n)\}$ with $\Pi_n$ the projector on the range of $(a-b)_+$; again
element $\id-\Pi_n$ is assigned to the null hypothesis $\rho^{\otimes n}$,
and element $\Pi_n$ is assigned to the alternative hypothesis $\sigma^{\otimes n}$.
We will show that this POVM asymptotically attains the Hoeffding bound.

Recall that inequality (\ref{eq:hay}) states
$$
\trace[a^s b^{1-s}] \ge \trace[\Pi b+(\id-\Pi)a].
$$
By positivity of $\trace[\Pi b]$ and $\trace[(\id-\Pi)a]$, this implies the two inequalities
$$
\trace[\Pi b],\trace[(\id-\Pi)a] \le \trace[a^s b^{1-s}].
$$
These yield the following upper bounds on the $\alpha$ and $\beta$ errors
of the chosen POVM (recall $Q_s=\trace[\rho^{1-s}\sigma^s]$):
\bea
\beta_n(\Pi_n) &=& \trace[(\id-\Pi_n)\sigma^{\otimes n}] \nonumber\\
&=& e^{nx}\trace[(\id-\Pi_n)a] \nonumber\\
&\le& e^{nx}\trace[a^s b^{1-s}] \nonumber\\
&=& e^{nx(1-s)}Q_s^n \nonumber\\
&=& \exp[n(x(1-s)+\log Q_s)].\label{eq:steina}
\eea
\bea
\alpha_n(\Pi_n) &=& \trace[\Pi_n \rho^{\otimes n}] \nonumber\\
&=& \trace[\Pi_n b] \nonumber\\
&\le& \trace[a^s b^{1-s}] \nonumber\\
&=& e^{-nxs}Q_s^n \nonumber\\
&=& \exp[n(-xs+\log Q_s)].\label{eq:steinb}
\eea
Choosing $x$ such that $x(1-s) +\log Q_s= -r$ then yields, from (\ref{eq:steina}),
\begin{eqnarray*}
\beta_n(\Pi_n) &\le& \exp(-nr),
\end{eqnarray*}
and from (\ref{eq:steinb}),
\begin{eqnarray*}
\alpha_n(\Pi_n) &\le& \exp\left(-n\left( -s\frac{r+\log Q_s}{1-s}-\log Q_s\right)\right) \\
&=&\exp\left(-n \frac{-rs -\log Q_s}{1-s}\right)\\&\leq& \exp\left(-n e_Q\left(r\right)\right),
\end{eqnarray*}
where  in the last inequality we have used the fact that the parameter $s$ was arbitrarily chosen from $(0,1)$.

Thus, for the rate limits we get
\begin{eqnarray*}
\beta_R\ge r,\quad \alpha_R\ge e_Q(r).
\end{eqnarray*}

The optimality, proven in the previous subsection, states that $\alpha_R\le e_Q(r)$ if $\beta_R=r$.
Furthermore, since $e_Q(r)$ is a non-increasing function, $\alpha_R \leq e_Q(r)$ if $\beta_R>r$.
This implies that for the chosen sequence of POVMs
$$
\beta_R = r,\quad \alpha_R = e_Q(r)
$$
must hold, which proves that the Hoeffding bound is indeed attained.
\qed
\subsection{Quantum Stein's Lemma and quantum version of Sanov's Theorem\label{sec:stein}}
The quantum generalisation of Stein's lemma deals with the asymptotics of the error quantity
\be\label{eq:alphastar}
\beta_n^*(\epsilon) := \inf_{\Pi_n} \{\beta_n(\Pi_n):\alpha_n(\Pi_n)\le \epsilon\},
\ee
for fixed $0<\epsilon<1$.
Here, the infimum is taken over all positive semi-definite contractions $\Pi_n$ on $\mathcal{H}^{\otimes n}$.

Quantum Stein's Lemma states that the rate limit $\beta_R^*(\varepsilon)$ of the sequence $(\beta_n^*(\epsilon))$
exists and is equal to $S(\rho \|\sigma)$, independently of $\epsilon$.
It was first obtained by Hiai and Petz \cite{hiaipetz}.
Its optimality part was then strengthened by Ogawa and Nagaoka in \cite{ogawa}.

Here we use the quantum HBCL Theorem to prove that the relative entropy $S(\rho \| \sigma)$ is
an achievable error rate limit and deduce optimality of this bound from Proposition 1 in \cite{qSanov2}.

\bigskip

\noindent\textit{Proof of the quantum Stein's lemma.}
We need to show that there is a sequence $\Pi$ with $\alpha(\Pi_n)\le \epsilon$ achieving
$\beta_R(\Pi)=S(\rho \|\sigma)$.
Let $\eta>0$ be small and set $r=S(\rho||\sigma)-\eta$.
 Achievability of the quantum Hoeffding bound means that a sequence $\Pi$ exists for which
$\beta_R\geq r$ and $\alpha_R=e_Q(r)$. Since $e_Q(r)>0$ for all $r < S(\rho \| \sigma)$ and $\eta>0$, the sequence $\alpha_n$ converges to 0.
Thus, from a certain value of $n$ onwards, $\alpha_n$ will get lower than any value $\epsilon>0$ chosen beforehand.
This means that $\Pi$ is a feasible sequence in (\ref{eq:alphastar}) for $n$ large enough, exhibiting
$\beta_R(\epsilon)\ge r=S(\rho \|\sigma)-\eta$.
As this holds for any $\eta>0$, we find that $\beta_R^*(\epsilon)\ge S(\rho \|\sigma)$.

With $\beta_R^*(\epsilon)\ge S(\rho \|\sigma)$ the two hypotheses associated to the pair of
density operators $(\rho, \sigma)$ satisfy the HP-condition in the terminology of the paper
\cite{qSanov2}.  Thus Proposition 1 in \cite{qSanov2} implies  $\beta_R^*(\epsilon) = S(\rho \| \sigma)$.\bigskip \qed

We remark that in \cite{qSanov2} the HP-condition was introduced for (ordered) pairs $(\Psi, \Phi)$
of arbitrary correlated states on quantum spin chains, while in the present paper only density
operators of the tensor-product form $\rho^{\otimes n}$ have been considered. These correspond
to the special case of shift-invariant product states on the infinite spin chain (quantum i.i.d.\ states).
A pair $(\Psi, \Phi)$  is said to satisfy the HP-condition if the relative entropy rate $s(\Psi \| \Phi)$
exists and is a lower bound on the lower rate limit $\underline \beta_R^*(\varepsilon)$ for all $\varepsilon \in (0,1)$.
\\ \\
Specifically to our setting (the i.i.d.\ case), Theorem 1 in \cite{qSanov2}  states that the achievability part in
quantum Stein's Lemma (the HP-condition) is equivalent to a quantum version of Sanov's theorem, which has
been presented in \cite{qSanov} and which is a priori
a result extending quantum Stein's Lemma in the following way:
\\ \\
Let the null hypothesis $H_0$ correspond to a family $\Gamma$ of density operators on $\mathcal{H}$ instead
of a single density operator $\rho$. Let the alternative hypothesis $H_1$ be still represented  by a fixed
density operator $\sigma$. Then there exists a sequence $\Pi$ of orthogonal projections $\Pi_n$ on
$\mathcal{H}^{\otimes n}$, respectively, such that for all $\rho \in \Gamma$ the corresponding type-I error
vanishes asymptotically, i.e.
\begin{eqnarray}\label{qSanov-constraint}
\lim_{n \to \infty} \trace [\rho^{\otimes n} \Pi_n]=0,
\end{eqnarray}
while the type-II error rate limit $\beta_R(\Pi)$ is equal to the relative entropy distance
from $\Gamma$ to $\sigma$:
$$
S(\Gamma \| \sigma) :=\inf_{\rho \in \Gamma} S(\rho \| \sigma).
$$
Moreover $S(\Gamma \| \sigma)$ is the upper bound on type-II error (upper) rate limit, for any sequence $\Pi$
of POVMs satisfying the constraint (\ref{qSanov-constraint}).
\\ \\
With the above reasoning we obtain the statement of quantum Sanov's Theorem from the quantum HBCL Theorem as well.
%
\section{Acknowledgements}
We thank the hospitality of various institutions: the Max Planck Institute for Quantum Optics (FV, KA),
the Erwin Schr\"odinger Institute in Vienna (FV, KA, AS), and the Physics Department of the National University of Singapore (KA).
KA was supported by The Leverhulme Trust (grant F/07 058/U), by the QIP-IRC (www.qipirc.org) supported by EPSRC
(GR/S82176/0), by EU Integrated Project QAP, and by the Institute of Mathematical Physics, Imperial College
London. MN has been supported by NSF under grant DMS-03-06497.
\appendix
\section{Proofs of Bounds on $Q$\label{appa}}
Inequality (\ref{eq:QFL}) stated in terms of general positive operators is
\begin{theorem}
For positive operators $A$ and $B$, and $0\le s\le 1$,
\be
\|A^{1/2} B^{1/2}\|_1 \le (\trace[A^{s} B^{(1-s)}])^{1/2} \, (\trace[A])^{(1-s)/2} \,(\trace[B])^{s/2}.
\ee
\end{theorem}
Specialising to states, $A=\sigma$ and $B=\rho$, the left-hand side is just $F(\rho,\sigma)$,
while the right-hand side is equal to $Q_s(\rho,\sigma)^{1/2}$.

\textit{Proof.}
We rewrite $A^{1/2} B^{1/2}$ as a product of three factors
$$
A^{1/2} B^{1/2} = A^{(1-s)/2} (A^{s/2}B^{(1-s)/2}) B^{s/2},
$$
apply H\"older's inequality on the 1-norm of this product, and exploit the
relation
$$
\|X^p\|_q = \|X\|^p_{pq}
$$
(for $X\ge0$) a number of times.
\beas
\|A^{1/2} B^{1/2}\|_1 &=& \|A^{(1-s)/2} (A^{s/2}B^{(1-s)/2}) B^{s/2}\|_1 \\
&\le& \|A^{(1-s)/2}\|_{2/(1-s)} \, \|A^{s/2}B^{(1-s)/2}\|_{2}\, \|B^{s/2}\|_{2/s} \\
&=& (\trace[A])^{(1-s)/2} \, \|A^{s/2}B^{(1-s)/2}\|_{2}\, (\trace[B])^{s/2} \\
&=& (\trace[A^{s}B^{(1-s)}])^{1/2} \, (\trace[A])^{(1-s)/2}(\trace[B])^{s/2}.
\eeas
\qed

\bigskip

We now give a direct proof of inequality (\ref{eq:QT}) that
circumvents the proof of (\ref{eq:TF}) and goes through
in infinite dimensions. We state it in terms of general positive operators:
\begin{theorem}
For positive operators $A$ and $B$,
\be
\|A-B\|_1^2 + 4(\trace[A^{1/2} B^{1/2}])^2 \le (\trace(A+B))^2.
\ee
\end{theorem}
\textit{Proof.} Consider two general operators $P$ and $Q$, and define their sum and difference as $S=P+Q$ and
$D=P-Q$. We thus have $P=(S+D)/2$ and $Q=(S-D)/2$. Consider the quantity
\beas
PP^*-QQ^* &=& \frac{1}{4}\left((S+D)(S+D)^* - (S-D)(S-D)^*\right) \\
&=& \frac{1}{2}(SD^*+DS^*).
\eeas
Its trace norm is bounded above as
\beas
\|SD^*+DS^*\|_1/2 &\le& (\|SD^*\|_1 + \|DS^*\|_1)/2 \\
&=& \|SD^*\|_1 \\
&\le& \|S\|_2 \|D\|_2.
\eeas
In the last line we have used a specific instance of H\"older's inequality for the
trace norm (\cite{bhatia} Cor.\ IV.2.6). Now put $P=A^{1/2}$ and $Q=B^{1/2}$, which exist by positivity of $A$ and
$B$, and which are themselves positive operators. We get $S,D=A^{1/2}\pm B^{1/2}$, hence
$$
\|A-B\|_1 \le \|A^{1/2} + B^{1/2}\|_2 \,\,\|A^{1/2} - B^{1/2}\|_2,
$$
which upon squaring becomes
\beas
\|A-B\|_1^2 &\le& \trace(A^{1/2} + B^{1/2})^2 \,\,\trace(A^{1/2} - B^{1/2})^2 \\
&=& \trace(A+B+A^{1/2}B^{1/2}+B^{1/2}A^{1/2})\\
&& \times \trace(A+B-A^{1/2}B^{1/2}-B^{1/2}A^{1/2}) \\
&=& (\trace(A+B)+2\trace(A^{1/2}B^{1/2}))\\
&& \times (\trace(A+B)-2\trace(A^{1/2}B^{1/2})) \\
&=& (\trace(A+B))^2 -4(\trace(A^{1/2}B^{1/2}))^2.
\eeas
\qed


\end{document}